\documentclass[12pt,preprint]{aastex}
\usepackage{amssymb}
\usepackage{graphicx}
\usepackage{lscape}
\usepackage{amsmath}
\usepackage{epsfig}

\begin{document}
\title{Multiband Fitting to Three Long GRBs with Fermi/LAT Data:
Structured Ejecta Sweeping up a Density-Jump Medium}
\author{S. Y. Feng$^{1,2}$, Z. G. Dai$^{1,2}$}
\affil{$^1$School of Astronomy and Space Sciences, Nanjing
University, Nanjing 210093, China;
\\$^2$Key Laboratory of Modern Astronomy and Astrophysics (Nanjing University),
Ministry of Education, Nanjing 210093, China;
\\siyifeng@msn.com. dzg@nju.edu.cn}

\begin{abstract}
We present broadband (radio, optical, X-ray and GeV) fits to the
afterglow light curves and spectra of three long-duration gamma-ray
bursts (GRBs 080916C, 090902B, and 090926A) detected by the
Gamma-Ray Burst Monitor (GBM) and Large Area Telescope (LAT)
instruments on the Fermi satellite. Using the observed broadband
data, we study the origin of the high energy emission, and suggest
that the early-time GeV emission and the late-time radio, optical,
and X-ray afterglows can be understood as being due to synchrotron
emission from an external forward shock caused by structured ejecta
propagating in a wind bubble jumping to a homogeneous density
medium. If the ceasing time for majority of the energy injection is
assumed to be close to the deceleration time of the forward shock,
the structured ejecta with continuous energy injection to the
forward shock can well explain the early rising feature of the GeV
mission from these burst, and the density-jump medium can account
for some certain plateaus or flares in the late afterglows. From our
fits, we find that, on one hand, the external shock origin of the
GeV photons will make the optical depth have not significant
contribution to the early LAT rising part, which will loosen strong
constraint of lower limits of Lorentz factor. On the other hand,
these Fermi-LAT events preferentially occur in a low-density
circumburst environment, in which case the Klein-Nishina cutoff will
significantly suppress the Self-Synchrotron Compton (SSC) radiation.
Such an environment might result from superbubbles or
low-metallicity progenitor stars (which have a low mass-loss rate at
late times of stellar evolution) of type Ib/c supernovae.
\end{abstract}

\section{Introduction}
\label{sect:intro}

Gamma-ray bursts (GRBs) are a kind of astrophysical phenomenon so
luminous in the universe that their isotropic energies of
$10^{48}-10^{55}{\rm ergs}$ are usually considered coming from
extremely relativistic outflows with bulk Lorentz factors as high as
$10^2-10^3$. The connection between long-duration GRBs and
broad-line SNe Ib/c (low-metallicity) has been supported by some
pieces of observational evidence \citep{Woosley06}. The recently
launched Fermi Gamma-Ray Space Telescope with the on-board Gamma-ray
Burst Monitor (GBM) and Large Area Telescope (LAT) instruments
(\citealt{Atwood09,Meegan09}) in conjunction with the \textit{Swift}
narrow field instruments \citep{Gehrels04} have opened a new era of
understanding physical mechanisms of GeV photon emission in very
energetic GRBs and their relation to lower-energy afterglow
emission. Up to now, several novel yet tricky features have appeared
during the whole period of observations and their complexities have
challenged the anterior established theoretic models.

The first feature is that the sub-MeV and GeV photons observed by
GBM and LAT respectively behave distinctive spectral and temporal
properties. The GBM light curves are nearly flat during the main
episode of the prompt emission, then drop extremely rapidly [e.g.
$t^{-3.3}$ for GRB 080916C \citep{Abdo09}], and eventually cease
abruptly. This can also be validated from $\sim60\%$ of all bursts
detected by the \textit{Swift} satellite \citep{Evans09}. For the
LAT light-curves, however, a rise appearing in the early few seconds
($<T_{90}$) was followed by a single power-law decay without any
cutoff up to hundreds of seconds after the GBM trigger, till below
the monitor sensitivity, and decay of the late LAT emission was much
shallower than the GBM-detected counterparts at the same times
\citep{Bzhang10}. Furthermore, the LAT emission usually lags the GBM
emission from a fraction of seconds to a few seconds. In addition,
the spectral slopes of the GBM and LAT emissions are often
different, e.g., the GBM data can be fitted with a Band spectrum
that is composed of two smoothly-joining power laws, while the LAT
data are often fitted by a power law with a slope intermediate
between the two slopes of the GBM fit. These properties seem to
indicate that the high-energy emission and low-energy emission
detected by LAT and GBM have different origins.

The temporal properties of the LAT emission have been studied, and
some explanations differing in the source's dominant component have
been proposed. A prevalent explanation is particle-dominated models.
One inclination is the leptonic interpretation. Because of their
distinctive light curve properties we mentioned above, it is highly
probable that the MeV photons may be of an internal origin, while
the GeV photons may be generated via synchrotron emission of
electrons accelerated by an external forward shock that also leads
to lower-energy afterglow emission \citep{Kumar09}.
\citet{Wang09,Wang10} studied the Klein-Nishina effect on the
high-energy afterglow emission and found that at early times such an
effect strongly suppress the inverse Compton scattering of those
electrons that produce the high-energy emission. Thus synchrotron
radiation of the electrons becomes a plausible mechanism. This
conclusion was independently drawn by \citet{zhang09}, who suggested
that the lack of a thermal component in the GBM spectrum of GRB
080916C is consistent with a relativistic Poynting-flux-dominated
outflow. The explanation of synchrotron radiation of the electrons
accelerated by a forward shock is fairly natural to account for both
the observed delay of the $>100$ MeV photons and long lasting of
their emission. However, a thermal component deviating from the
smooth Band spectrum function in some bursts (i.e. GRB090510,
GRB090902B, GRB090926A) seems to be beyond a prediction of what is
often invoked from the external shock model \citep{Bzhang10}.
Moreover, a rapid rise during the first few seconds (e.g. $\sim t^6$
of GRB080916C) is hard to be well explained within such a framework
\citep{Toma09}. Therefore, in addition to the hypothesis of a
separate origin, \citet{Toma09} assumed that GeV and MeV photons may
come from the same region, but the onset of the high-energy may
result from anisotropic inverse Compton scattering of an
optical-thin expanding cocoon, delayed compared with the MeV
emission. Nevertheless, in their calculation, this up-scattering
cocoon is so short-lived that it could not account for the whole
high energy emission.

Another approach is the hadronic scenario. \citet{Razzaque09}
suggested that the MeV and GeV photons could be interpreted as the
radiation of accelerated electrons and cosmic ray protons
respectively, as well as the delay between them could come from
different cooling time scales in a highly magnetized shock. In their
framework, the model of GRB 080916C is plausible only when
$\Gamma\le500$ and the jet opening angle $\sim1^{\circ}$.
Alternatively, \citet{Asano09} considered another possibility of the
photomeson cascade and proton synchrotron models, and provided their
constraint in GRB090510. Notwithstanding, to reproduce the extra
component around GeV with these models, the isotropic-equivalent
proton injection luminosity is required to be larger than
$10^{55}{\rm erg\,s^{-1}}$. Such a large proton luminosity is a
challenge for the hadronic models.

All the works mentioned above inspire us to consider a plausible
structured outflow, in which the bulk Lorentz factor of the initial
shells tends to be lower than that of the late shells. This energy
accumulation, therefore, would lead to an early rapidly rising light
curve and transient soft to hard spectrum.

The second feature of the three long GRBs with LAT data is that some
humps or flares appear in the light curves of the $>10^5$ s
low-energy afterglows (\citealt{Cenko10,Swenson10}). Neglecting this
feature, \citet{Kumar10} fitted the multi-band light curves of these
GRBs by assuming that a relativistic external shock sweeps up an
interstellar medium, and obtained reasonable physical parameters or
parameter spaces accordingly. Although their fittings somehow favor
the external shock model, humps (or sometimes flares) have indeed
been observed on the late-time optical and X-ray afterglow light
curves, which do not completely accord with a simple power-law but
instead the decay after the humps is shallower than that before the
humps. These observations call for a more meticulous consideration
of the external shock model.

It is noted that a density-jump medium proposed by \citet{Dai02} has
provided us with a clue for some optical and X-ray humps and well
fitted to several bursts (\citealt{Dai03,Tam05,Jin09}). In this
scenario, a relativistic jet first expands in a stellar wind,
subsequently encounters a density jump, and finally expands in a
homogeneous medium.  This interaction can produce an observed
light-curve bump.

In this paper, we show that the early-time GeV emissions together
with the late-time radio, optical, and X-ray afterglows of GRBs
080916C, 090902B, and 090926A can be understood as being due to
synchrotron emission from an external forward shock caused by
structured ejecta propagating in a wind bubble jumping to a
homogeneous density medium.  In Sec.\ref{sect:obs} we include a set
of observed broadband (LAT, XRT, UVOT) data on these three luminous
long bursts. Then, our model is set up in Sec.\ref{sect:model}. In
particular, The structured ejecta can well explain the universal
early rising feature of the GeV emission from these bursts, and the
density-jump medium can account for some certain plateaus and flares
in the late afterglows. Additionally, for the sake of verifying
whether or not the lower energy (X-ray, optical, radio) emission
originates from the same source as the higher energy ($>$100 MeV)
emission, we discuss the effect of synchrotron self-Compton
(synchrotron self-absorption) on the high energy (radio) emission,
which, in the constrained parameter space estimated analytically
from XRT and UVOT light curves, is proved in Sec.\ref{sect:fit} to
have a small contribution to the flux density during the observed
period. In this section, we find a reasonable set of parameters
valid for most of the late afterglows. Our conclusions concerning a
plausible central engine and ambient environment of bright, long
GRBs are discussed in Sec.\ref{sect:con}.

Throughout this work, we adopt the convenience $Q_x=Q/10^x$ in units
of cgs.

\section{Observations}
\label{sect:obs}

 Among the 19 observed Fermi-LAT GRBs during the
first 2 years' operation, GRB080916C, GRB090902B and GRB090926A are
3 typical brightest long GRBs with abundant of spectra information
\citep{Granot10}.

\subsection{GRB080916C}

This burst, located at redshift $z=4.35\pm0.15$ \citep{Greiner09},
is the first GRB detected by Fermi with high significance of photons
at energies $>0.1$ GeV. The isotropic energy emitted from prompt
emission can be estimated as $E_{\gamma, iso} =8.8\times10^{54}{\rm
erg }$ \citep{Abdo09}. At 00:12:45.613542 on 16 September 2008
\citep{Abdo09}, this GRB was triggered on by GBM with the duration
$T_{90}=66$s. Before $T_0+6$s, the LAT light-curve shows an
extremely steep rise $F_{\nu,LAT}\propto t^{6\pm0.5}$, followed by a
simple power law decay $F_{\nu,LAT}\propto
t^{-1.33\pm0.08}\nu^{-1.1\pm0.1}$ until $\sim 1400$s
\citep{Bzhang10}. X-ray and optical photons were detected by XRT and
UVOT since 17:11:28 16 September 2008 ($T_0+61$ks). A steep decay
(to $\sim T_0+101$ks) continued with a plateau (to $\sim T_0+204$ks)
goes to a slightly shallower decay without break until 1.3Ms from
the trigger $T_0$. During the period of $T_0+61$ks$-T_0+1306$ks, the
simple power law light curve and spectrum evolution show the flux
density at X-ray band, $F_{\nu,X}\propto
t^{-1.29\pm0.09}\nu^{-0.50\pm0.16}$, as well as for the optical
band, $F_{\nu,opt}\propto t^{-1.40\pm0.05}\nu^{-0.38\pm0.20}$
\citep{Greiner09}. Besides, AGILE, RHESSO, INTEGRAL, Konus-Wind, and
MESSENGER all provide a plentiful of data information for the late
afterglow \citep{Perri08}.

Many groups have studied this GRB with the external shock model
(\citealt{Kumar10,Gao09,Zou09}), and suggested that the late time
afterglow data can be used to extrapolate the early LAT data as well
as be predicted from it, under the preference of circumburst density
stratification.

\subsection{GRB090902B}
This burst is an exceptional case with redshift $z=1.822$, whose
speciality lied in both phases of prompt emission and early
afterglow \citep{Abdo09b}. Unlike majority of the other GRB events,
excess emission exhibited in both low ($\le100$keV) and high
($\ge10$MeV) band during the prompt emission. Moreover, the
soft-hard-soft spectral evolution indicates two components: Band
function peaking at $\sim700$keV + simple power-law with photon
index $\Gamma\sim1.85$ \citep{Bzhang10}. The isotropic energy
release $E_{\gamma,iso}=(3.83\pm0.05)\times10^{54}{\rm erg }$
\citep{Cenko10}. There are some explanations for origin of the
high-energy power-law emission component observed in the LAT energy
range (which accounts for $\sim24\%$ of the total 10keV to 10GeV
fluence), including the hadronic origin [either proton synchrotron
radiation \citep{Razzaque09} or photohadronic interactions
\citep{Asano09}], or thermal emission from the jet photosphere
\citep{Ryde10}.

This burst was triggered and located at 11:05:08.31 on September
2009, by the Fermi-GBM \citep{Bissaldi09} and Suzaku-WAM
\citep{Terada09} with multipeaked duration $T_{90}\approx 21s$.
After a rapid rise until $T_0+7$s, the LAT band light-curve decays
up to $T_{0}+1$ks with $F_{\nu,LAT}\propto t^{-1.4\pm0.06}$,
including an energetic photon detected as high as
$11.16^{+1.48}_{-0.58}$ GeV within the prompt emission phase and
another $33.4^{+3.7}_{-3.5}$ GeV at $T_0+82$s \citep{Abdo09b}. The
\textit{Swift} XRT (\citealt{Kennea09,Evans09b,Stratta09}) and UVOT
\citep{Swenson09} began concurrently target of opportunity of the
field of GRB090902B's fading source from 23:36 on 2 September 2009
and copious of data have been in hand from $T_0+0.5$ks. A steep
decay (to $\sim T_0+116$ks) simultaneously ended with a slight rise
in both bands. It is highly possible that jet break occurred at
$\sim T_0+553$ks \citep{Cenko10}. During this whole period, the
X-ray spectrum evolved as $F_{\nu,X}\propto
t^{-1.36\pm0.03}\nu^{-0.90\pm0.13}$, and for the optical band,
$F_{\nu,opt}\propto t^{-0.89\pm0.05}\nu^{-0.76\pm0.07}$. In
addition, VLA began to observe the afterglow since 3 September 2009
\citep{Chandra09} until 5 months later at 8.5GHz and 4.8GHz.

Using the forward-reverse shock and constant density model,
\citet{Cenko10} suggested the afterglow of this burst is better
fitted in X-ray and radio bands than in the optical/UV band, except
for the first point at $\sim10^4$s. Parameter constraints without
taking LAT data into account suggest a low circumburst density and a
large kinetic energy, which is possible in the narrow opening angle
$\sim3.4^{\circ}$. Later, \citet{Liu10} have provided a meticulous
discussion about the two-component forward-reverse external shock in
a monotonic circumburst environment and their calculation was well
fit to the four observed bands, except for the very early rising
part of the LAT light curve.

\subsection{GRB090926A}
This burst is the GRB detected at $z=2.1062$ by Fermi with photons
as high as $\sim 20$ GeV at 26s after the trigger \citep{Uehara09}.
The prompt emission time-integrated flux indicates its isotropic
emission energy of $E_{\gamma,
iso}=2.1_{-0.08}^{+0.09}\times10^{54}{\rm erg }$ \citep{Bzhang10}.

Since the trigger at 04:20:26:26.99 on 26 September 2009, GBM and
LAT started operation and found that multiple pulses with total
duration $T_{90}\approx 20s$ can be well modeled by a Band function
spectrum in the prompt emission, along with the Suzaku-WAM
\citep{Noda09}, Konus-Wind \citep{Golenetskii09}, and RT-2 on
CORONAS-PHOTON \citep{Chakrabarti09}. The light curve rose slightly
for quite a few seconds (to $\sim T_0+16$s), and decayed
$F_{\nu,LAT}\propto t^{-2.05\pm0.14}\nu^{-1.26_{-0.22} ^{+0.24}}$. A
fading X-ray counterpart was observed by XRT since 17:17 on 26
September ($T_0+46.7$ks) \citep{Vetere09}, then PROMPT, SMARTS
detected photons in several optical filters but no radio (5.5GHz)
source under the limit of 1.5mJy was detected up to 1 October 2009.
As shown on the light curve, a plateau ($T_0+51.4$ks $\sim
T_0+92$ks) overlaps the simple power law decay that is slightly
steeper than the fading supposed to be extrapolated from the early
trace. No break up to 100ks from the trigger $T_0$ was observed.
During the period from $T_0+46.7$ks to $T_0+149$ks, two
variabilities, are suggested to be flares (the first, at $\sim70{\rm
ks}-95{\rm ks}$ with $\delta t/t\approx0.35$; the second is slight,
at 195ks-260ks, with $\delta t/t\approx0.28$) overly on the simple
power law light curve in both X-ray band ($F_{\nu,X}\propto
t^{-1.40\pm0.05}\nu^{-1.6_{-0.2} ^{+0.3}}$)and optical band
($F_{\nu,opt}\propto t^{-1.01_{-0.07} ^{+0.03}}$) \citep{Swenson10}.

\citet{Cenko10} and \citet{Rau10} studied the later afterglow
(X-ray, optical) of this burst, provided a certain parameter space,
confirmed the second flare, and indicated that the jet break occurs
around $\sim21$d. However, due to lack of radio data, the
constraints cannot be narrowed down. Meanwhile, as for a common
rebrightening plateau in 5 optical bands, they suggested a
density-jump circumstance (e.g., \citealt{Dai02};
\citealt{Lazzati02,Tam05}) or a smooth injection of energy into the
forward shock from the central engine
(e.g.,\citealt{Dai98a,Rees98}), and called for a detailed analysis
for the early higher energy emission.

\section{Model}
\label{sect:model}
 The behavior of the GRB afterglows shed light on
the external shock model \citep{Panaitescu02}. A collimated
($\theta\le10^{\circ}$), ultrarelativistic outflow of matter and/or
radiation is driven by a central engine. Some certain dissipative
processes within the outflow give rise to the prompt gamma-ray
emission, with a fraction
$\varsigma_{\gamma}\equiv{E_{\gamma,iso}/(E_{\gamma,iso}+E_{K,iso})}$
of the total relativistic energy converted to high-energy radiation.

A self-consistent result of the multiband (X-ray, optical, radio)
afterglow data coordinated with the prompt emission relics producing
$>100{\rm MeV}$ photons, may support the external shock origin of
LAT photons, which is the same as the later lower-band emission.
Meanwhile, due to the difference between the light curve slope of
LAT and GBM, we take no consideration of the GBM emission in this
piece. Moreover, the particularities of GRB 090902B we mentioned
above (soft-hard-soft spectral evolution, flares) may evoke doubts
about the conventional external shock + monotonic circumstance model
(we will discuss it in Sec.\ref{sect:con}), but enlighten our
deliberate consideration on amendment to structured ejecta sweeping
up the density-jump medium.

\subsection{Dynamics}
\label{sect:dyna}
\subsubsection{Before the Deceleration Time}
Energy can be injected by a central engine continuously during a
period (majority of which is within the first main pulse of prompt
emission, before $t_{fp}$\footnote[1]{As Fig. 2, 10, 11 in
\citet{Bzhang10} shown, majority of the $>100$ MeV photons in the
prompt emission are released during the first largest pulse (at
$\sim10\rm s$,$\sim10\rm s$,$\sim10\rm s$ for GRB 080916C, GRB
090902B, and GRB 090926A, separately). Comparing this time to the
peaking time of the LAT lightcurve ($\sim6\rm s$,$\sim7\rm
s$,$\sim16\rm s$ for GRB 080916C, GRB 090902B, and GRB 090926A,
separately), and the $T_{90}$ ($\sim66\rm s$,$\sim21\rm
s$,$\sim20\rm s$ for GRB 080916C, GRB 090902B, and GRB 090926A,
separately), we defined $t_{fp}$, before when energy has been
largely injected. Of course, as mentioned in \citet{Maxham10}, whole
of the energy injection last longer, while it will dwindle after the
first largest pulse. }). We here suggest, within the framework of
the collapsar model, an increase of the Lorentz factor of the ejecta
in active time of the central engine may due to two reasons. First,
angular momentum of the accreted fall-back matter would spin up the
central compact object and thus its rotational energy loss could
give rise to an increase of the ejecting luminosity. Second,
earlier-ejected shells, when breaking through the stellar envelope,
may suffer from more massive baryon contamination, and thus
later-ejected shells may propagate in a tunnel (the later the
cleaner). Therefore, the Lorentz factor of the ejecta head may
increase with time. For simplicity, we assume the bulk Lorentz
factor of the front materials distribute as a power-law function of
time: $\Gamma\propto {t}^\varkappa$ for $t\le t_{fp}$. If
$\varkappa>0$, the Lorentz factor of the shocked matter increases,
being due to energy injection.

We denote $M_{\rm ej}$ as the accumulated mass ejected by the
central engine at some certain time, and $\eta$ as the bulk Lorentz
factor of a blast wave at this certain time, which is defined when
the mass of the surrounding matter swept up by the blast wave,
$M_{\rm sw}$, is equal to $M_{\rm ej}/\eta$. After this time, the
bulk Lorentz factor presents an obvious deceleration, contributing
to a peak of the LAT light curve. This time is called the
deceleration time $t_{dec}$, which is assumed in this paper to be
around the stop time of effective energy injection. This assumption
is reasonable because energy injection to the forward shock is
effected by the circumburst medium. Thus, both the apparent
energy-injection time $t_{fp}$ and the deceleration time $t_{dec}$
should be shorter than the central engine ceasing time $T_{90}$.

The shock propagates a distance $\delta R\sim 2(1+z)^{-1}\Gamma^2
c\delta t$ during a small observed time $\delta t$. Assuming the
proton number density of a medium $n\propto R^{-k}$, thus for a
homogeneous medium ($k=0$), $n={\rm const}$ (\citealt{Sari98}) or
for a wind medium ($k=2$), $n=AR^{-2}=10^{35.5}A_{\rm 35.5}R^{-2}$
(\citealt{Dai98b,Chevalier00}). In both cases, we obtain
\begin{equation}\label{tdec}
t_{\rm dec}=\frac{(2\varkappa+1)(1+z)}{2 \eta^2 c} R_{\rm dec}.
\end{equation}
The deceleration radius can be written as
\begin{equation}\label{radius}
R_{\rm dec}={\left(\frac{3-k}{4\pi A m_p } \frac{E_{\rm
K,iso}}{\eta^2 c^2}\right)}^{1/(3-k)}.
\end{equation}

Therefore, $R(t\le t_{dec})=R_{\rm dec}(t/t_{\rm
dec})^{2\varkappa+1}$. According to \citet{Sari97}, in order to
check whether the the unshocked ejecta is assumed to be equal to
that of the shocked matter, let $\triangle=c t_{fp}(1+z)^{-1}$
represent the thickness of the shell, and $f$ be the density ratio
between the preshock fluid in the shell and in the circumburst
surrounding ($k=2$), given by
\begin{equation}\label{f}
f= \frac{(1+\varkappa)(1+z)E_{\rm {K,iso}}}{4 \pi A R^{2-k} m_p
c^3{\eta}^2}{\frac{{t_{fp}}^{\varkappa+1}}{t^{\varkappa+2}}},~~~~~~~~~~~~~~~~~t\le
t_{fp}
\end{equation}
This may bring about a correction for the GeV rising part under
different conditions, i.e., in the thick shell case, $f\ll\Gamma^2$,
a reverse shock is relativistic, the Lorentz factor of the shocked
fluid should be changed and the contribution of the reverse shock to
the flux density should be considered, while in the thin shell case,
$f\succsim \Gamma^2$, this change is negligible. However, no matter
how the thickness of the shell is, $t_{fp}\le t_{dec}$. In the
following analytical calculations, we simplify that $t_{fp}\sim
t_{dec}$ (which can be observed from the figures of temporal flux
density) and assume that the self-similar condition is established
after the deceleration time.

 Before the
deceleration time, energy is continuously injecting into the former
shells. The mass ejected later catches up with the earlier ones, and
$\Gamma$ can be simplified as the bulk Lorentz factor of these
materials combination. In the very beginning, it is possible that
the bulk Lorentz factor is too low for majority of high-energy gamma
rays to escape from the MeV background. According to predominate
hypothesis (\citealt{Granot10}), only when the optical depth of the
$\gamma$-ray absorption $\tau_{\gamma\gamma}<1$, the bulk Lorentz
factor is large enough to radiate almost completely. In this piece,
we use sketchy in \citet{Zou11} to check whether the absorption play
a dominate role in certain parameter spaces.

\subsubsection{In the Density-Jump Surrounding}
Because of the association of long GRBs with star forming regions,
it is highly possible that massive stars are still embedded in a
cloud when giving birth to GRBs. Low density bubbles are created by
stellar winds from GRB progenitors, whose sizes and densities
strongly depend on the initial ambient density. Therefore, a density
jump occurs at the boundary between the wind bubble and the outer
cloud, which may result in a bump/flare and shallow decay of the
light curve during the later period \citep{Tam05}.

From the moment $t_{\rm break}$ at which the blast wave reaches the
boundary, to the termination when it comes out into the interstellar
medium at $t_{\rm end}$, the number density of the medium changes as
\begin{equation}
n=A R^{-k}=\left\{\begin{array}{l} 3\times 10^{35} A_{35.5} R^{-2},
~~~~~~~~~~~~~~~~~~~~~~~~~~~~~~~~~~~~~~~~~~~R\le R_{\rm break},\\
3\times 10^{35}[1+\frac{{\log}R-{\log}R_{\rm break}}{{\log}R_{\rm
end}-{\log}R_{\rm break}}(\xi-1)]A_{35.5} R_{\rm break}^{-2},
~~~~~R_{\rm break}\le R\le R_{\rm
end},\\
n_{\rm after}\equiv 3\times 10^{35}\xi A_{35.5} R_{\rm break}^{-2},
~~~~~~~~~~~~~~~~~~~~~~~~~~~~R\ge R_{\rm end},
\end{array}\right.
\end{equation}
where $\xi$ is the ratio of the number density in front of and
behind the jump.

It has been usually assumed that the afterglow emission from the
blast wave is nearly adiabatic (\citealt{Meszaros97,Sari98}), so the
total kinetic energy of the relativistic shock is constant. However,
if we take radiative energy loss into consideration, the radiation
efficiency of the blast wave can be given with the electronic
fraction $\epsilon_e$ by \citet{Wu05}
\begin{equation} \label{loss}
\varepsilon=\left\{\begin{array}{l} \epsilon_e,
~~~~~~~~~~~~~~~~~~~~\nu_c\le
\nu_m,\\(\frac{\nu_m}{\nu_c})^{(p-2)/2}, ~~~~~~~\nu_m\le \nu_c.
\end{array}\right.
\end{equation}
In such a situation, energy loss is significant in the fast cooling
and gradually dwindle to the quasi-adiabatic case (i.e., the slow
cooling phase).

Hereafter, a denotation $m=(3-k)/(1-\varepsilon)$ can analytically
result in the bulk Lorentz factor evolution,
\begin{equation}
\Gamma=\left\{\begin{array}{l} {\eta(\frac{t}{t_{\rm
dec}})}^\varkappa, ~~~~~~~~~~~~~~~~~~~~~~~~ t\le t_{\rm dec},\\\eta
{(\frac{t}{t_{\rm dec}})}^{-m/(2m+2)}, ~~~~~~~~~~~~~t\ge t_{\rm
dec}.
\end{array}\right.
\end{equation}
Consequently, the Doppler factor becomes
$\mathcal{D}$$=\frac{1}{\Gamma (1-\beta cos
\theta)}=\frac{2\Gamma}{1+\Gamma^2\theta^2}\sim 2\Gamma$ (where
$\theta$ is the emitting latitude angle).

When the blast wave reaches the density boundary, the observed light
curve will not behave as an abrupt jump but a smooth rise or plateau
instead, due to the curvature effect (\citealt{Fenimore96,Kumar00}).
Although this is trifle to be accomplished in analytic modeling, a
simplified power law distribution of medium density in the period of
$t_{\rm break}\sim t_{\rm end}$ can account for the same light curve
feature (See Fig.~\ref{fig:parameter}.a). Succeedingly, the blast
wave radius can be written as
\begin{equation}
R=\left\{\begin{array}{l} R_{\rm dec}{(\frac{t}{t_{\rm
dec}})}^{2\varkappa+1}, ~~~~~~~~~~~~~~~~~t\le t_{\rm dec},\\
R_{\rm dec}{(\frac{t}{t_{\rm
dec}})}^{(1-\varepsilon)/(4-k-\varepsilon)}, ~~~~~~~~t\ge t_{\rm
dec}.
\end{array}\right.
\end{equation}

\subsection{Radiation}
\label{sect:rad}
 In the comoving frame, with an electronic
fraction $\epsilon_e$ and magnetic fraction $\epsilon_B$ of the
post-shock thermal energy density, which may either vary or not with
different media. Denote $f_p\equiv 6(p-2)/(p-1)$, the magnetic field
intensity $B=\Gamma ({32 \pi \epsilon_Bn m_p c^2})^{1/2}$. The
cooling, minimum, and maximum electron Lorentz factors are
$\gamma_{c}=\gamma_{syn,c}/[1+Y(\gamma_c)]=6\pi m_ec
(1+z)(\sigma_TB^2\Gamma t)^{-1}/[1+Y(\gamma_c)]$,
$\gamma_m=(1/6)f_p{\epsilon_e}(m_p/m_e)\Gamma$, and
$\gamma_{\max}\sim 10^8/\sqrt{B}$, while the synchrotron
characteristic frequencies $\nu_c$, $\nu_{m}$, $\nu_{max}$ can be
calculated by $\nu_i=\gamma_{i}^2\mathcal{D}eB/[2\pi m_ec (1+z)]$
(where $i=c, m, \max$), where $Y(\gamma_*)$ is the Compton factor
for electrons with the Lorentz factor $\gamma_*$, defined as the
ratio of the self-inverse Compton (SSC) to synchrotron emission
(\citealt{Nakar09}).
\begin{equation}\label{Y}
Y(\gamma_*)\equiv {P_{SSC}(\gamma_*) \over {P_{syn}(\gamma_*)}}
\end{equation}
 When Klein-Nishina (KN) effects are unimportant, i.e., the IC scattering
of $\gamma_*$ electrons with synchrotron photons are in the Thomson
scattering regime, $Y(\gamma_*)$ can be simply derived by
\citep{Sari01} as $(\sqrt{1+4\varepsilon
\epsilon_e/\epsilon_B}-1)/2$, which is independent with $\gamma_*$.
However, for high-energy electrons whose KN effect becomes
important, $Y(\gamma_*)$ depends on $\gamma_*$ and can be expressed
as analytical solutions in different cases from \citet{Wang10}.

The total number of shocked accelerated electrons $N_e=4\pi n
R^3/(3-k)$. The peak spectral power of a single electron is
$P_{\nu,\max}$$=\sigma_T m_ec^2 \mathcal{D}\mathit{B}/(3e)$, and the
peak flux density at certain time of the afterglow is
$F_{\nu,\max}=(1+z)N_e P_{\nu,\max}/(4\pi{D_L}^2)$. Here, the
luminosity distance $D_L=(1+z){H_0}^{-1}c\int\limits_0^z
[\Omega_m(1+z)^3+\Omega_k(1+z)^2+\Omega_{\Lambda}]^{-1/2}dz$ (here
we adopt a standard ${\rm \Lambda CDM}$ cosmology with $H_0=71{\rm
kms^{-1}Mpc^{-1}}$, $\Omega_m=0.27$, $\Omega_k=0.0$,
$\Omega_{\Lambda}=0.73$ \citep{Spergel07}).

For a lower energy band (e.g., radio), synchrotron self-absorption
may play a crucial role. The self-absorption optical depth
$\tau_{\nu,m}$ or ($\tau_{\nu,c})$, according to the definition
$\tau_{\nu}=1$, becomes
$\tau_{\nu,m}=\frac{c_0(p-1)}{(3-k)}\frac{enR}{B{\gamma_m}^5} $ and
$ \tau_{\nu,c}=\frac{c_0}{(3-k)}\frac{enR}{B{\gamma_c}^5}$, where
$c_0\approx10.4(p+2)/(p+2/3)$ \citep{Wu05}. Succeedingly, the
self-absorption frequency $\nu_{a}$ can be written as
\begin{equation}
\nu_{a}=\left\{\begin{array}{l} \tau_{\nu,l}^{3/5}\nu_l,
~~~~~~~~~~~~~~~~~~~~~~~\tau_{\nu,l}<1,\\
\tau_{\nu,l}^{2/(p+4)}\nu_l,
~~~~~~~~~~~~~~~~~~1\le\tau_{\nu,l}<({\nu_h\over\nu_l})^{(p+4)/2},\\
\tau_{\nu,l}^{2/(p+5)}({\nu_h\over{\nu_l}})^{1/(p+5)}\nu_l,
~~~~~({\nu_h\over\nu_l})^{(p+4)/2}\le\tau_{\nu,l}<({\nu_{\max}\over{\nu_l}})^{(p+5)/2}({\nu_l\over{\nu_h}})^{1/2}.
\end{array}\right.
\end{equation}
Then, the SSA power-law light curve yields \citep{Sari01}
\begin{equation} \label{syn}
F_{\nu}=F_{\nu,max}\left\{\begin{array}{l}
({\nu_a\over{\nu_l}})^{1/3}({\nu\over{\nu_a}})^2, ~~~~~~~~~~~~~~~~~~\nu\le \nu_a,\\
({\nu\over{\nu_l}})^{1/3}, ~~~~~~~~~~~~~~~~~~~~~~~~\nu\le \nu_l,\\
({\nu\over{\nu_l}})^{-(q-1)/2}, ~~~~~~~~~~~~~~~~~~\nu_l\le\nu\le \nu_h,\\
({\nu_h\over{\nu_l}})^{-(q-1)/2}({\nu\over{\nu_h}})^{-p/2},
~~~~~~~~\nu\ge\nu_h.
\end{array}\right.
\end{equation}
Here $\nu_l={\rm min}(\nu_m,\nu_c)$, $\nu_h={\rm max}(\nu_m,\nu_c)$.
$q=2$ when $\nu_c<\nu_m$, while $\nu_c>\nu_m$, $q=p$.

It is noted that before the optical depth becomes thin, the
annihilation effect of high-energy photons can affect their flux, so
the observed flux can be estimated as $F_{\rm
ob,\nu}=e^{-\tau_{\gamma\gamma}}F_{\rm \nu}$. After $T_{90}$,
$F_{\rm ob,\nu}=F_{\rm \nu}$.

For a high energy band, synchrotron self-Compton (SSC) radiation
might contribute to the observed flux, the characteristic
frequencies $\nu_{IC,c}$, $\nu_{IC,m}$, $\nu_{IC,\max}$,
$F_{IC,\nu,\max}$ can be calculated by $\nu_{IC,i}=2\nu_
i\gamma_{i}^2$ (where $i=c,m,\max)$, and
$F_{IC,\nu,\max}=(14/45)\sigma_T RnF_{\nu,\max}$. Thus the SSC
power-law light curve yields \citep{Gupta07}
\begin{equation}\label{ssc}
F_{\nu,IC}={F_{IC,\nu, \max}}\left\{\begin{array}{l}
({\nu\over{\nu_{IC,l}}})^{1/3}, ~~~~~~~~~~~~~~~~~~~~~~~~~~~~~~~~~~~~~~~~~~~~~~~~~~~\nu\le \nu_{IC,l},\\
({\nu\over{\nu_{IC,l}}})^{-(q-1)/2}, ~~~~~~~~~~~~~~~~~~~~~~~~~~~~~~~~~~~~~~~~~~~~~\nu_{IC,l}\le\nu\le\nu_{IC,h},\\
({\nu_{IC,h}\over{\nu_{IC,l}}})^{-(q-1)/2}({\nu\over{\nu_{IC,h}}})^{-p/2}, ~~~~~~~~~~~~~~~~~~~~~~~~~~~~~~~\nu_{IC,h}\le\nu
\end{array}\right.
\end{equation}
Here $\nu_{IC,l}={\rm min}(\nu_{IC,m}$, $\nu_{IC,c})$,
$\nu_{IC,h}={\rm max}(\nu_{IC,m}$, $\nu_{IC,c})$. $q=2$ when
$\nu_{IC,c}<\nu_{IC,m}$, while $\nu_{IC,c}>\nu_{IC,m}$, $q=p$.

However, due to the KN effects on optically thin spectrums,
Equations \eqref{syn} and  \eqref{ssc} should be amended as
different cases mentioned in equations of \citet{Nakar09}. This
calls for the definition of electron KN Lorentz factor:
$\widehat{\gamma_i}=m_e c^2 \Gamma/[h \nu_{syn}(\gamma_i)]$
($i=c,m$). We have, if the synchrotron photons emitted by electrons
with Lorentz factor larger than $\gamma_i$ [i.e. with frequency
$>\nu_{syn}(\gamma_i)]$, they cannot be upscattered efficiently by
electrons with Lorentz factor larger than $\widehat{\gamma_i}$
(which is above the KN limit).

In the case of a jet, a simple assumption is that the jet
decelerates without sideways expansion. After the jet break time
$t_{jet}$ at which the jet angle $\theta_j\sim 1/\Gamma$ (where the
jet axis is assumed to be along the line of sight), the observed
flux $F_{\nu,\rm jet}=F_{\nu}({t/ t_{jet}})^{(3-k)/(4-k)}$.

Comparing the flux density estimated analytically above with the
observational multi-band values, certain physical parameters can be
constrained. Whether the parameter space is reasonable or not can on
some extent verify the feasibility of our external shock model.

\section{Fitting Results}
\label{sect:fit}
\subsection{Constraints on Parameters }
Based on the analytical model we mentioned above, our procedure of
further narrowing down the physical parameter space in the
self-similar phase includes the following steps:

In the wind environment:

(1) According to the expression of $F_{\nu}\propto
t^{-\alpha}\nu^{-\beta}$ in different spectral regimes given by
\citet{Wu05} and the observed high-energy, X-ray and optical
temporal and spectral indices, $(p,\varepsilon)$ can be roughly
estimated (see Tab.~\ref{tab:data}).

(2) For simplicity, $\varepsilon\approx\epsilon_e$ in the fast
cooling regime and then $\varepsilon$ decays slowly (where we denote
$t_{cm}$ as the time when $\nu_c= \nu_m$, and obtain that
$\varepsilon$ decays as $\varepsilon\sim\epsilon_e (t/t_{cm})^{2-p}$
in the slow cooling regime (i.e., setting $j\equiv -m/(2m+2)$, we
have, after $t_{cm}$, $\Gamma\propto t^j$, then $\nu_m\propto
t^{-(j+1)}$ and $\nu_c\propto t^{(j+1)}$, and equation \eqref{loss}
provides $\varepsilon\propto\epsilon_e(\nu_m/\nu_c)^{(p-2)/2}\propto
t^{(2-p)(j+1)}$, as long as $\varepsilon\le 2/3$, $j+1\sim 2/3$).

(3) Because of the definition for the deceleration time [set $k=2$
in Equations \eqref{tdec} and \eqref{radius}], $\eta$  can be
expressed with $\varkappa$, $E_{\rm K,iso}$ and $n$ (or $A_{\rm
35.5}$). This can estimate whether the shell thickness is thin
enough to neglect the reverse shock's contribution to the total flux
density.

(4) Hereafter the free parameters are $\varkappa$, ${E}_{\rm
K,iso}$, $\epsilon_{\rm B}$, and $n$ (or $A_{\rm 35.5}$). The
observed flux density in different wavebands (radio, optical, X-ray,
$>$100 MeV) can be combined and eventually simultaneously be solved.
In this paper, the temporal indices of the X-ray and optical light
curves concurrently evolve according to equation \eqref{syn} and
thus $\epsilon_B$ can therefore be expressed with $\varkappa$, $
{E}_{\rm K,iso}$ and $n$ (or $A_{\rm 35.5}$).

(5) To justify whether the early high energy LAT photons come from
the same region as the late afterglow, what we need to do is to
replace $\epsilon_B$ in the form of $\varkappa$, ${E}_{\rm K,iso}$
and $n$ (or $A$) from step (4). To confirm whether SSC make a
significant contribution to the GeV afterglow or Klein-Nishina
effect changes the synchrotron and SSC spectra, we need to discuss
the value of $Y(\gamma_c)$ in the slow cooling regime or
$Y(\gamma_m)$ in the fast cooling regime \citep{Nakar09}, therefore
$Y(\gamma_{GeV})$ can be estimated \citep{Wang10}. If no solution is
found in the range of ($10^{53}\,{\rm ergs}\precsim  {E}_{\rm
K,iso}\precsim 10^{56}\,{\rm ergs}$ and $10^{-2}\precsim A_{35.5}
\precsim 10^{2}$), the $>$100 MeV photons may come from the other
component, vise vesa.

(6) From discussions about the parameter space of ($A_{35.5}$ and
$\varkappa$) according to observational statistics, one can obtain
several best fittings. In addition, The ratio $\xi$ of the wind and
homogeneous medium density, $\epsilon_e$ and $\epsilon_B$ at early
and late times may have slight changes, which can also be legally
adjusted in practise.

\subsection{Best Fittings}
We put the above procedure into operation, and best fitting
parameters are shown in Tab.~\ref{tab:fitting}. The evolution of
characteristic frequencies for each burst is expected as follows.
Here our LAT data are taken from \citet{Bzhang10} and X-ray data
were reported by \textit{Swift} (\citealt{Evans07,Evans09}).

From the lightcurves, we found that from $t_{dec}$ on,
$\alpha_{GeV,X,opt}>0$, $\beta_{GeV,X,opt}>0$. According to the
analytical expression from \citet{Wu05}, before the blast wave
reaches the density jump, GeV photons at early decaying time (e.g.,
$t_{dec}\le t=t_I\le 10^3$s) has two possibilities:
\begin{eqnarray*}
F_{syn,GeV}\propto \left\{\begin{array}{l}
{t}^{-\frac{3p-2-(p-2)\varepsilon}{2(2-\varepsilon)}\nu_{GeV}^{-p\over 2}}, ~~~~~~~~~~~~~~~~~~~~~~~~~~\nu_{GeV}\ge\nu_{m}(t)\ge\nu_{c}(t),\\
{t}^{-\frac{3p-2-(p-2)\varepsilon}{2(2-\varepsilon)}+\frac{p-2}{4-p}}\nu_{GeV}^{-p\over
2}, ~~~~~~~~~~~~~~~~~~~\nu_{GeV}\ge\nu_{c}(t)\ge\nu_{m}(t).
\end{array}\right.
\end{eqnarray*}
 while the lower energy bands (X-ray, optical) at
$t=t_{II}>10^{4} $s, may lie in one of the following slow cooling
cases:
\begin{eqnarray*}
 F_{opt}\propto
{t}^{-\frac{3p-1-(p-1)\varepsilon}{2(2-\varepsilon)}}\nu_{opt}^{-(p-1)\over
2}, F_{X}\propto
{t}^{-\frac{3p-1-(p-1)\varepsilon}{2(2-\varepsilon)}}\nu_X^{-(p-1)\over
2},~~~~~~~~~~~\nu_{c}(t)\ge \nu_X\ge \nu_{opt} \ge \nu_{m}(t),\\
F_{opt}\propto
{t}^{-\frac{3p-1-(p-1)\varepsilon}{2(2-\varepsilon)}}\nu_{opt}^{-(p-1)\over
2}, F_{X}\propto
{t}^{-\frac{3p-2-(p-2)\varepsilon}{2(2-\varepsilon)}+\frac{p-2}{4-p}}\nu_X^{-p\over
2},~~~~~~~~~~\nu_X\ge \nu_{c}(t)\ge \nu_{opt} \ge \nu_{m}(t),
\\
F_{opt}\propto
{t}^{-\frac{3p-2-(p-2)\varepsilon}{2(2-\varepsilon)}+\frac{p-2}{4-p}}\nu_{opt}^{-p\over
2}, F_{X}\propto
{t}^{-\frac{3p-2-(p-2)\varepsilon}{2(2-\varepsilon)}+\frac{p-2}{4-p}}\nu_X^{-p\over
2}, ~~~~~~~~\nu_X\ge \nu_{opt}\ge \nu_{c}(t) \ge \nu_{m}(t).
\end{eqnarray*}
for 090902B radio band, at $t=t_{III}$ when $\beta_{radio}>0$
(i.e.$\nu_{radio} \ge \nu_{m}$), it is similar to the optical
spectrum and lightcurve. While $\beta_{radio}<0$,
\begin{eqnarray*}
F_{radio}\propto{t}^{-\frac{\varepsilon}{3(2-\varepsilon)}}\nu_{radio}^{1\over
3}, ~~~~~~~~~~~~~~ \nu_{m}(t) \ge \nu_{radio}.
\end{eqnarray*}

\subsubsection{GRB080916C}
Our optical photometry of this burst is taken from
\citet{Greiner09}. First of all, if the X-ray and optical afterglow
shares the same source as the high-energy photons detected by LAT,
their spectral indices $\beta_{ob,x}=0.50\pm0.16$ and
$\beta_{ob,opt}=0.38\pm0.02$ should be consistent with the GeV
photons, $\alpha_{ob,LAT}=1.33\pm0.08$ and
$\beta_{ob,LAT}=1.1\pm0.1$. This is preliminarily valid only in the
case of $\nu_{GeV}\ge {\rm max}[\nu_{c}(t_{I}),\nu_{m}(t_{I})]$ and
$\nu_{c}(t_{II})\ge \nu_X\ge \nu_{opt} \ge \nu_{m}(t_{II})$,when
$2.0\le p\le 2.32$ and $0<\epsilon_e\le 0.48$.

Consequently, the peak of the LAT light curve lying at
$t_{dec}\sim6$s can be considered as the deceleration time of this
burst. Therefore, equations \eqref{tdec} and \eqref{radius} lead to
the relationship among the isotropic kinetic energy $E_{\rm K,55}$,
the bulk Lorentz factor $\eta_{3}$, and the
 wind parameter $A_{35.5}$ at the very moment: $\eta_3\sim0.40E_{\rm K,55}^{1/4}
 A_{35.5}^{-1/4}(1+2\varkappa)^{1/4}$.

Equation \eqref{f} provides, at $t\le t_{dec}$,
${f/{\Gamma^2}}=72\exp(5.37\varkappa)(1+\varkappa)(1+2\varkappa)^{-1}t^{-3\varkappa-2}>1$.
Thus the reverse shock is nearly Newtonian.

Subsequently, when the blast wave is embedded in the wind bubble, we
assume that the typical value $\varepsilon(t_I)\sim 1/3$, while
$\varepsilon(t_{II})\sim 0.0$, and the characteristic frequencies
evolve as
\begin{eqnarray}
\nu_{c}(t_I)=1.21\times 10^9\,{\rm Hz} E_{\rm K,55}^{1/2}
A_{35.5}^{-2} \epsilon_{B,0}^{-3/2}(1+2\varkappa)^{-5/2}{t_I}^{2/5}[1+Y(\gamma_c,t_{I})]^{-2}, \\
\nu_{c}(t_{II})=1.03\times 10^9\,{\rm Hz} E_{\rm K,55}^{1/2}
A_{35.5}^{-2}
\epsilon_{B,0}^{-3/2}(1+2\varkappa)^{-5/2}{t_{II}}^{1/2}[1+Y(\gamma_c,t_{II})]^{-2},\\
\nu_{m}(t_I)=8.66\times 10^{23}\,{\rm Hz} E_{\rm K,55}^{1/2}
 \epsilon_{e,0}^{2} \epsilon_{B,0}^{1/2}(1+2\varkappa)^{3/2}{t_I}^{-8/5}{f_p}^{2}, \\
\nu_{m}(t_{II})=7.39\times 10^{23}\,{\rm Hz} E_{\rm K,55}^{1/2}
 \epsilon_{e,0}^{2} \epsilon_{B,0}^{1/2}(1+2\varkappa)^{3/2}{t_{II}}^{-3/2}{f_p}^{2}.
\end{eqnarray}

Setting $p=2.3$ and $\epsilon_{e,0}=1/3$ as a trail, we only
consider the synchrotron emissivity,
\begin{equation}\label{080916GeV}
F_{syn,{\rm GeV}}(13 {\rm s})=18.5\mu{\rm Jy}E_{\rm
K,55}^{43/40}\epsilon_{B,0}^{3/40}(1+2\varkappa)^{9/40}[1+Y(\gamma_c,
13 \rm s)]^{-1}.
\end{equation}
We found that equation \eqref{080916GeV} is approximately the same
under both fast cooling and slow cooling condition, so do with the
following equation \eqref{090902GeV} and \eqref{090926GeV}.
\begin{equation}\label{080916syn}
F_{syn,X}(1.01\times 10^5 {\rm s})=1.76\times 10^5\mu{\rm Jy}E_{\rm
K,55}^{33/40}A_{35.5}\epsilon_{B,0}^{33/40}(1+2\varkappa)^{59/40}\sim
7.73\times 10^{-2}\,\mu {\rm Jy}.
\end{equation}
From  equation \eqref{080916syn}, we have $\epsilon_{B,0}=1.96\times
10^{-8}E_{\rm K,55}^{-1}A_{35.5}^{-40/33}(1+2\varkappa)^{-9/5}$, so
$t_{cm}\le 13$s in the case that $10^{-2}\precsim A_{35.5}$.
Therefore, $\nu_{m}(13 \rm s)<\nu_{c}(13 {\rm s})$ and $F_{syn,{\rm
GeV}}(13 \rm s)\le 0.83\mu{\rm Jy}$, which yields
\begin{equation}
5.83E_{\rm K,55}A_{35.5}^{-1/10}(1+2\varkappa)^{1/10}\le
1+Y(\gamma_c, 13{\rm s})<54E_{\rm
K,55}A_{35.5}^{1/5}(1+2\varkappa)^{-1/5}
\end{equation}
From \citet{Wang10}, we have, in the wind, if $t_I>13{\rm s}$,
$Y(\gamma_c,t_I)<Y(\gamma_c,13 {\rm s})$. This have been shown in
Fig.~\ref{fig:parameter}.c. Hence, as long as
$A_{35.5}(1+2\varkappa)^{-1}{t_I}<20.2$ (e.g.$10^{-2}\precsim
A_{35.5}(1+2\varkappa)^{-1}<1.56$ at 13s or $10^{-2}\precsim
A_{35.5}(1+2\varkappa)^{-1}\sim0.02$ at $10^3$ s),
\begin{equation}\label{080916Y}
{\widehat{\gamma_{m}(t_I)}\over \gamma_{c}(t_I)}=3.0\times
10^{-3}E_{\rm
K,55}^{-1}A_{35.5}^{2/5}{t_I}^{3/5}(1+2\varkappa)^{-2/5}[1+Y(\gamma_c,t_I)]<0.18A_{35.5}^{3/5}{t_I}^{3/5}(1+2\varkappa)^{-3/5}<1,
\end{equation}

According to \citet{Nakar09}, in the slow cooling, equation
\eqref{080916Y} leads to $Y(\gamma_c,t_I)<1$, then SSC cooling has
no effect on the electron distribution. From the analytical solution
of \citet{Wang10}, $Y(\gamma_{GeV},t_I)\ll$1 can be estimated. With
the assumption that the GeV luminosity is dominated by synchrotron
emission,$F_{syn,{\rm GeV}}(13 \rm s)\sim 0.83\mu{\rm Jy}$ provide a
stronger constraint, $E_{\rm
K,55}=0.19A_{35.5}^{1/10}(1+2\varkappa)^{-1/10}$.In such parameter
space, we have $\tau_{\gamma\gamma}\ll$1, so the absorption does not
contribute to the early rising.

All the above relations show that $\epsilon_{e,0}$, $E_{K,55}$ and
$A_{35.5}$ can be expressed with $\eta_3$ and $\varkappa$, the
physical mechanism for the rising part of the lightcurve can be
considered as contribution of the structured ejecta. According to
observational statistics, the best fitting yields $E_{\rm
K,55}=0.13$, $A_{35.5}=0.02$, $\epsilon_{e,0}=0.3$,
$\epsilon_{B,0}=5.81\times10^{-6}$, and $\eta_3=0.74$. As the result
shown in Fig.~{fig:080916C}: before the deceleration time, the bulk
Lorentz factor of the swept blast wave increases with
$\varkappa=0.5$ (which is self-consistent with our assumption for
$\varkappa$ and thin shell case), $t_{cm}$ occurs at $\sim4.5$s then
electrons are slow cooling, and deceleration begins from $\sim6$ s,
the density jump is expected around $10^{5.3}- 10^{5.6}$ s (i.e.,
$3.25\times10^{18}$ cm). After breaking out to the interstellar
medium, the surrounding density is around $1.17\times10^{-3} {\rm
cm^{-3}}$.

\subsubsection{GRB090902B}
Along with the  LAT and XRT data mentioned above, our optical
photometry here is taken  from \citet{Pandey10}, as well as the
8.5GHz data reported by the VLA \citep{van der Horst09}. With the
assumption that the late X-ray, optical and radio afterglow and the
early LAT photons originate from the same region, their spectral
indices $\beta_{ob,x}=0.90\pm0.13$ and $\beta_{ob,opt}=0.76\pm0.07$
should be consistent with the GeV photons,
$\alpha_{ob,LAT}=1.4\pm0.06$. The same as the above analysis, $\nu_{
GeV}\ge {\rm max}[\nu_{c}(t_{I}),\nu_{m}(t_{I})]$ and
$\nu_{c}(t_{II})\ge \nu_X\ge \nu_{opt} \ge \nu_{m}(t_{II})$,
respectively, which result in the plausible light curve and spectrum
similar to the above case. Therefore, roughly speaking, $2.3\le p\le
2.6$ and $0.17\le\epsilon_e\le 0.43$ are the available region.

Although the peculiar ``soft-hard'' spectral evolution seems tricky
in the prompt emission phase, its duration is so short ($T_0-T_0+10$
s) and it ends before the deceleration time \citep{Bzhang10}, which,
is understandable in the frame of structured ejecta acceleration (we
will discuss it in Sec.\ref{sect:con}), so the external shock model
is still applicable. In addition, although a 33 GeV photon was
detected at 82 s after the trigger, we consider the 11.2 GeV photon
at $\sim10$\,s as the highest energy photon of estimation for the
lower Lorentz factor because the arrival time of the 33GeV photon is
far beyond the prompt emission phase and may not merely be
attributed from the external shock but some other origins.

The turning point of the LAT light curve occurs at $\sim$7 s. This
indicates that a relationship among $E_{\rm K,55}$, $\eta_{3}$, and
$A_{35.5}$ becomes $\eta_3\sim0.33E_{\rm
K,55}^{1/4}A_{35.5}^{-1/4}(1+2\varkappa)^{1/4}$.
 Hence, equation \eqref{f} yields that at $t\le t_{dec}$ and
 ${f/{\Gamma^2}}=98\exp(5.94\varkappa)(1+\varkappa)(1+2\varkappa)^{-1}
t^{-3\varkappa-2}>1$, so the thin shell case is applicable in this
fitting.

Similar to analytical solutions of GRB080916C we mentioned above,
characteristic frequencies of the photons observed by LAT and later
multiband (X-ray, optical, radio) afterglow can be estimated.
Additionally, when we come to the radio band, SSA may be crucial.

Here again, we assume that the typical value $\varepsilon(t_I)\sim
1/3$, while $\varepsilon(t_{II})\sim 0.0$, with model parameters
($E_{\rm K,55},\epsilon_{B,0},\epsilon_{e,0},$ and $A_{35.5}$), the
characteristic frequencies evolve as
\begin{eqnarray}
\nu_{c}(t_I)=3.29\times 10^{9}\,{\rm Hz}E_{\rm K,55}^{1/2}
A_{35.5}^{-2} \epsilon_{B,0}^{-3/2}(1+2\varkappa)^{-5/2}{t_I}^{2/5}[1+Y(\gamma_c,t_{I})]^{-2}, \\
\nu_{c}(t_{II})=2.71\times 10^{9}\,{\rm Hz} E_{\rm K,55}^{1/2}
A_{35.5}^{-2}
\epsilon_{B,0}^{-3/2}(1+2\varkappa)^{-5/2}{t_{II}}^{1/2}[1+Y(\gamma_c,t_{II})]^{-2},\\
\nu_{m}(t_I)=6.52\times 10^{23}\,{\rm Hz} E_{\rm K,55}^{1/2}
 \epsilon_{e,0}^{2} \epsilon_{B,0}^{1/2}(1+2\varkappa)^{3/2}{t_I}^{-8/5}{f_p}^{2}, \\
\nu_{m}(t_{II})=5.37\times 10^{23}\,{\rm Hz} E_{\rm K,55}^{1/2}
 \epsilon_{e,0}^{2} \epsilon_{B,0}^{1/2}(1+2\varkappa)^{3/2}{t_{II}}^{-3/2}{f_p}^{2}.
\end{eqnarray}

Setting $p=2.4, \epsilon_{e,0}=0.4$ as a trail, the synchrotron
emissivity is given by
\begin{equation}\label{090902GeV}
F_{syn,{\rm GeV}}(10 \rm s)=14.58\mu{\rm Jy}E_{\rm
K,55}^{11/10}\epsilon_{B,0}^{1/10}(1+2\varkappa)^{3/10}[1+Y(\gamma_c,10
\rm s)]^{-1}.
\end{equation}
\begin{equation}\label{090902syn}
F_{syn,X}(1.17\times 10^5 \rm s)=6.72\times 10^5\mu{\rm Jy}E_{\rm
K,55}^{17/20}A_{35.5}\epsilon_{B,0}^{17/20}(1+2\varkappa)^{31/20}\sim
0.18\mu {\rm Jy}
\end{equation}
From equation \eqref{090902syn}, we have $\epsilon_{B,0}=1.87\times
10^{-8}E_{\rm K,55}^{-1}A_{35.5}^{-20/17}(1+2\varkappa)^{-9/5}$, so
$t_{cm}\le 10$s in the case that $A_{35.5}\ge 10^{-2}$. Therefore,
$\nu_{m}(10 \rm s)<\nu_{c}(10 \rm s)$ and $F_{syn,{\rm GeV}}(10 \rm
s)\le 1.59\mu{\rm Jy}$, which yields,
\begin{equation}
1.54E_{\rm K,55}A_{35.5}^{-3/25}(1+2\varkappa)^{3/25}\le
1+Y(\gamma_c,10 \rm s)<55E_{\rm
K,55}A_{35.5}^{9/50}(1+2\varkappa)^{-9/50}.
\end{equation}
Meanwhile, for the radio band at $t_{III}$ s, because of the light
curve index $\sim 1/3$, SSA can be neglected, we have
\begin{eqnarray}
\nu_{m}(5.23\times 10^5 \rm s)=9.14 \times
10^{10}A_{35.5}^{-10/17}(1+2\varkappa)^{10/17}>8.5 {\rm GHz},\\
F_{syn,8.5{\rm GHz}}(5.23\times 10^5 \rm s)=1.89 \times 10^2\mu{\rm
Jy}A_{35.5}^{3/5}(1+2\varkappa)^{-3/5}\sim 56_{-26}^{+24}\mu{\rm
Jy}.
\end{eqnarray}
Therefore, it should be $0.05<A_{35.5}(1+2\varkappa)^{-1}<0.24$, and
if $A_{35.5}(1+2\varkappa)^{-1}{t_I}<46.7$
(e.g.$A_{35.5}(1+2\varkappa)^{-1}<0.24$ at 10s or
$0.05<A_{35.5}(1+2\varkappa)^{-1}\sim 0.05$ at $10^3$ s),
\begin{equation}\label{090902Y}
{\widehat{\gamma_{m}(t_I)}\over \gamma_{c}(t_I)}=1.8\times
10^{-3}E_{\rm
K,55}^{-1}A_{35.5}^{2/5}(1+2\varkappa)^{-2/5}t^{3/5}[1+Y(\gamma_c,t_I)]<0.10A_{35.5}^{3/5}(1+2\varkappa)^{-3/5}{t_I}^{3/5}<1.
\end{equation}

Under such a parameter space, equation \eqref{090902Y} leads to
$Y_{c,t_I}<1$, SSC makes a little effect because of the
Klein-Nishina suppression, so the estimation that
$Y(\gamma_{GeV},t_I)\ll$1 is self-consistent. With the assumption
that the GeV luminosity is dominated by synchrotron emission,
$F_{syn,{\rm GeV}}(10 \rm s)\sim 1.59\mu{\rm Jy}$ provide a stronger
constraint, $E_{\rm K,55}=0.65A_{35.5}^{2/17}(1+2\varkappa)^{-2/17}$
and $\tau_{\gamma\gamma}\ll1$.

Using the above relations, we can express $\epsilon_{e,0}$,
$E_{K,55}$ and $A_{35.5}$ by $\eta_3$ and $\varkappa$, and fit the
data according to observational statistic. The best fitting requires
$E_{\rm K,55}=0.44$, $A_{35.5}=0.06$, $\epsilon_{e,0}=0.4$,
$\epsilon_{B,0}=4.56\times 10^{-7}$, and $\eta_3=0.60$.
Fig.~{fig:parameter} shows the circumburst density, radius and
characteristic frequency, and Compton $Y_{\rm GeV}$ and
$Y_{\gamma_c}$ factor evolution under such a parameter space.
Fig.~{fig:090902B} shows the spectrum and light curve fitting.
Before the deceleration time (7s), the bulk Lorentz factor of the
blast wave increases with a slope of $\varkappa=0.8$ (which is
self-consistent with $\varkappa$ and thin shell case). The
shock-accelerated electrons are cooling fast early before $\sim0.72$
s, the density jump is estimated around $10^{5.1}-10^{6}$ s (i.e.,
at radius of $4.48\times10^{18}$ cm). After breaking out to the
interstellar medium, the surrounding density is as low as
$1.91\times10^{-3}\,{\rm cm^{-3}}$. These parameters are consistent
with the radio light curve except for the first data and is reliable
when the jet break takes place nearly at the same time. We find that
the jet angle is 0.04 rad and collimation-corrected energy is
$E_{jet}\sim1.51\times10^{51}\, {\rm ergs}$.

\subsubsection{GRB090926A}
V band and R band data of this burst are taken from
\citet{Swenson10} and \citet{Rau10}, respectively. Our analysis
procedure is the same as mentioned above. Due to the hypothesis of
the same source, the temporal index of the LAT photons
$\alpha_{ob,LAT}=2.05\pm0.14$, should be consistent with the
spectral indices $\beta_{ob,x}=1.12\pm0.13$ and
$\beta_{ob,opt}=1.03\pm0.05$. This leads to a fairly broad plausible
range for $3\ge p \ge2$ and $\epsilon_e\sim 10^{-1}$ (e.g.under the
condition of $\nu_{GeV}>{\rm max}[\nu_{m}(t_I),\nu_{c}(t_{I})]$,
$\nu_{c}(t_{II})>\nu_{R,V}$ requires for $3\ge p\ge 2.96$ and
$0.62>\varepsilon>0.37$, while $2.5\ge p\ge 2$ and
$0.70>\varepsilon>0.45$ as long as $\nu_{c}(t_{II})<\nu_{X}$).
However, since the light curve after the density jump is a little
bit steeper than the early part, $\varepsilon$ (the latter the
smaller) should not deviate too much from the typical value (i.e.,
$1/3$, otherwise, the decrease of loss efficiency will cause the
decaying slope much shallower), whereas the decay of LAT is steep,
which requires a large $p$ (i.e., $3.0$). Besides, in the afterglow
phase, we can reasonably believe that the first flare at $70-95$ ks
comes from an interaction of the blast wave and density jump. The
second ``flare'' may come from late energy injection or another
density jump, and in this paper, we smooth it as the simple power
law.

Because the deceleration time is estimated as $\sim16$ s, $\eta_{3}$
can be written by $\eta_3\sim0.27E_{\rm
K,55}^{1/4}A_{35.5}^{-1/4}(1+2\varkappa)^{1/4}$. Hence, at $t\le
t_{dec}$, ${f/
{\Gamma^2}}=512\exp(8.32\varkappa)(1+\varkappa)(1+2\varkappa)^{-1}
t^{-3\varkappa-2}>1$, so the reverse shock is negligible in this
fitting. With the typical value $\varepsilon(t_I)\sim 1/3$, while
$\varepsilon({t_{II}})\sim 0.0$, The characteristic frequencies
evolve as
\begin{eqnarray}
\nu_{c}(t_I)=3.09\times 10^{9}\,{\rm Hz} E_{\rm K,55}^{1/2}
A_{35.5}^{-2} \epsilon_{B,0}^{-3/2}(1+2\varkappa)^{-5/2}{t_I}^{2/5}[1+Y(\gamma_c,t_{I})]^{-2}, \\
\nu_{c}(t_{II})=2.34\times 10^{9}\,{\rm Hz} E_{\rm K,55}^{1/2}
A_{35.5}^{-2}
\epsilon_{B,0}^{-3/2}(1+2\varkappa)^{-5/2}{t_{II}}^{1/2}[1+Y(\gamma_c,t_{II})]^{-2},\\
\nu_{m}(t_I)=7.43\times 10^{23}\,{\rm Hz} E_{\rm K,55}^{1/2}
 \epsilon_{e,0}^{2} \epsilon_{B,0}^{1/2}(1+2\varkappa)^{3/2}{t_I}^{-8/5}{f_p}^{2}, \\
\nu_{m}(t_{II})=5.63\times 10^{23}\,{\rm Hz} E_{\rm K,55}^{1/2}
 \epsilon_{e,0}^{2} \epsilon_{B,0}^{1/2}(1+2\varkappa)^{3/2}{t_{II}}^{-3/2}{f_p}^{2}.
\end{eqnarray}

Setting $p=3.0, \epsilon_{e,0}=0.4$ as a trail, the synchrotron
emissivity is calculated by
\begin{equation}\label{090926GeV}
F_{syn,{\rm GeV}}(36 \rm s)=28.7\mu{\rm Jy}E_{\rm
K,55}^{5/4}\epsilon_{B,0}^{1/4}(1+2\varkappa)^{3/4}[1+Y(\gamma_c,36
\rm s)]^{-1},
\end{equation}
\begin{equation}\label{090926syn}
F_{syn,X}(5.3\times 10^4 \rm s)=1.42\times 10^6\mu{\rm Jy}E_{\rm
K,55}A_{35.5}\epsilon_{B,0}(1+2\varkappa)^{2}\sim 0.33\mu {\rm Jy}.
\end{equation}
From equation \eqref{090926syn}, we have $\epsilon_{B,0}=2.32\times
10^{-7}E_{\rm K,55}^{-1}A_{35.5}^{-1}(1+2\varkappa)^{-2}$, so
$t_{cm}=4.31/E_{\rm K,55}\le 36$s in the case that $E_{K,55}\ge
0.12$. Therefore, $\nu_{m}(36 \rm s)<\nu_{c}(36 \rm s)$ and
$F_{syn,{\rm GeV}}(36 \rm s)\le 0.37\mu{\rm Jy}$, which yields,
\begin{equation}
1.71E_{\rm K,55}A_{35.5}^{-1/4}(1+2\varkappa)^{1/4}\le
1+Y(\gamma_c,36 \rm s)<8.38E_{\rm K,55}.
\end{equation}

Hence,as long as
$A_{35.5}^{1/2}(1+2\varkappa)^{-1/2}{t_I}^{3/5}<65.8$ (e.g.,
$A_{35.5}(1+2\varkappa)^{-1}<58.7$ at 36s or
$A_{35.5}(1+2\varkappa)^{-1}<1$ at $10^3$ s),
\begin{equation}\label{090926Y}
{\widehat{\gamma_{m}(t_I)}\over \gamma_{c}(t_I)}=1.8\times
10^{-3}E_{\rm
K,55}^{-1}A_{35.5}^{1/2}(1+2\varkappa)^{-1/2}t^{3/5}[1+Y(\gamma_c,t_I)]
<0.015A_{35.5}^{1/2}(1+2\varkappa)^{-1/2}{t_I}^{3/5}<1.
\end{equation}

In the slow cooling case, equation \eqref{090926Y} leads to
$Y(\gamma_c,t_I)<1$, SSC makes a little effect due to the
Klein-Nishina suppression, because the estimation of
$Y(\gamma_{GeV},t_I)\ll$1 is self-consistent.

Meanwhile, during $1\le t_I\le 10^{2.5}$ s, in the case of fast
cooling ($E_{\rm 55}<0.12$) that $\nu_{m}(36 \rm s)>\nu_{c}(36 \rm
s)$ and $F_{syn,{\rm GeV}}(36 \rm s)\le 0.37\mu{\rm Jy}$, $
A_{35.5}(1+2\varkappa)^{1/5}\le 0.026$,
\begin{eqnarray}
{\gamma_{m}(t_I)\over \widehat{\gamma_{m}(t_I)}}=2.38\times
10^{3}A_{35.5}^{-1/2}(1+2\varkappa)^{1/2}{t_I}^{-8/5}>1,
\end{eqnarray}
According to \citet{Nakar09}, this is within the strong KN regime
(case II). Analytical discussion with the above confines yield to
the possibilities of case IIb and IIc, both leads to
$Y(\gamma_{GeV},t_I)\ll$1. Hence, the contribution from SSC could be
neglected, too. In both fast and slow cooling, $F_{syn,{\rm GeV}}(36
\rm s)\sim 0.37\mu{\rm Jy}$ provide a stronger constraint, $E_{\rm
K,55}=0.70A_{35.5}^{1/4}(1+2\varkappa)^{-1/4}$ and
$\tau_{\gamma\gamma}\ll1$.

Using the above relations, we can express $\epsilon_{e,0}$,
$E_{K,55}$ and $A_{35.5}$ by $\eta_3$ and $\varkappa$, and fit the
data according to observational statistic. When $E_{\rm K,55}=0.43$,
$A_{35.5}=0.03$,$\epsilon_{e,0}=0.35$,
$\epsilon_{B,0}=8.18\times10^{-6}$, and $\eta_3=0.56$, fitting seems
plausible (Fig.~{fig:090926A}). Before the deceleration time, the
Lorentz factor of the blast wave increases with a slope of
$\varkappa=0.3$, $t_{cm}\sim8.8$ s and from $\sim16s$ significant
deceleration appears. The blast wave enters the ISM around
$10^{4.8}-10^{4.9}$ s (i.e., at radius of $4.41\times10^{18}$ cm).
The medium density surrounding the bubble is $1.3\times10^{-3}\,{\rm
cm^{-3}}$.

\subsection{Analysis of Results}
Tab.~\ref{tab:fitting} shows all the best fit parameters for the
above three GRBs. Common results can be found:

(1) In all the three burst fittings, the thin shell case is
applicable, the very beginning of GeV flux densities are
forward-shock dominated. The rising slope comes from the
contribution of the power-law increasing bulk Lorentz factor of
structured ejecta. These are self-consistent with the thin shell
case, indicating that the central engine is very likely to be the
core collapse of a massive star (we will discuss it in
Sec.\ref{sect:con}). Our model implies that the central object would
keep on accreting matter even after the LAT trigger.

(2) The remaining isotropic kinetic energy of the fireball after the
prompt emission are all $\sim 10^{54}$ ergs, indicating $\sim
30\%-90\%$ of the total energy has been emitted in the prompt
emission. This is because $t_{cm}$ occurs before the deceleration
time for all the three GRBs and the following dynamics could be
considered as being quasi-adiabatic. As a matter of fact,
$\epsilon_e$ in our sample fittings is $\leqslant 2/3$, which meets
the requirement of radiation efficiency related to the analytic
procedure from \citet{Wu05}. Practically, slow cooling of electrons
starts before the deceleration time, and a jet break appears much
later, so Wu's spherical-like solution is tenable in this piece. A
low value of $\epsilon_B$ ($\sim 10^{-7}-10^{-5}$) indicates a weak
magnetic field, especially for GRB 080916C and GRB 090902B. These
GRBs have been studied by \citet{Kumar09}, whose inferred magnetic
field (along with several other LAT-detected events) is consistent
with shock compression of a modest circumstellar field ($B\succsim
30\mu $G). That is to say, no dynamo process is necessary to
generate the magnetic field needed for the observed synchrotron
afterglow emission. In our best fittings, we find $B>380\mu $G for
GRB080916C and $B>106\mu $G for GRB090902B even when $T_0+10^7$s,
being broadly consistent with the results of \citet{Piran10} and
\citet{Li10}. Meanwhile, when the blast wave enters the
constant-density circumburst medium, the preshock magnetic field can
be calculated with our fitting parameters as $B=(2\pi m_p \epsilon_B
n)^{1/2}\approx$ $8.0\mu $G for 080916C and $2.6\mu $G for 090902B,
that suggests no magnetic field amplification.

(3) Although the ratio of $\epsilon_e$ and $\epsilon_B$ is large,
synchrotron self-Compton is not significantly contributed to the
observed flux density in the GeV band because of the Klein-Nishina
suppression effect. In our sample fittings, only synchrotron
radiation from the external forward shock is enough to contribute to
the high energy emission and the synchrotron self-Compton component
can be neglected because $Y(\gamma_{GeV})\ll 1$ in this piece. In
particular, for GRB 090902B, from our best-fitting parameters, the
maximum electron Lorentz factor at 82s indicates the highest photon
energy can be as high as $\sim37$ GeV, so it is plausible that 33
GeV photon may result from synchrotron radiation.

(4) The wind parameter for three longest bursts is $A_{35.5}\sim
10^{\rm -2}-10^{\rm -1}$, which is reliably low. According to
\citet{Dai03}, the progenitor star of GRB 080916C, GRB 090902B, and
GRB 090926A could have been in a cloud. Due to a high pressure of
the cloud, a speedy wind will be slowed down by a pair of shocks
(viz, a reverse shock that propagates into the wind gas and a
forward shock that propagates into the cloud). This interaction
produces a stellar wind bubble and forms a density-jump at a radius
of $\sim10^{\rm 18}$ cm. The homogeneous density outside this jump
is estimated to be in the range of $n\sim10^{\rm -3}{\rm cm}^{-3}$,
which is much less than the typical density of an interstellar
medium, but could not be excluded from the possibility of existence
in areas of active star formation as the interiors of a prexisting
superbubble (\citealt{Panaitescu01,Scalo01}). If confirmed by the
superbubble observation, this can provide a piece of evidence for a
connection between long-duration GRBs and broad-lined SNe Ib/c. and
would be natural to expect that massive progenitors of GRBs explode
where we observe star formation. Meanwhile, in other fitting works
(\citealt{Cenko10,Liu10}), they have also confronted with the
problem of low density, and suggest the lower metallicity
progenitors with minimal pre-explosion mass loss or selection
effects \citep{Cenko10}. Hence, although the LAT emission at least
during $T_{90}$, from the spectral perspective \citep{Bzhang10}, is
likely to connect to the GBM emission of internal origin, the
fitting facts tentatively suggest a possibility that the LAT
observation records an internal-external shock transition and at
least in the LAT decaying phase the external forward shock emission
is dominated.

\section{Conclusions and discussions}
\label{sect:con}
 It is widely believed that long duration
gamma-ray bursts (GRBs), like hydrogen-deficient Type Ib/c
supernovae (SNe Ib/c), result from the core collapse of a massive
star. The main characteristic of setting GRBs apart from other SNe
is that a substantial fraction of the explosive energy is coupled to
relativistic ejecta. A compact central engine is responsible for
accelerating and collimating a jet-like outflow and driving a SN
explosion (\citealt{Woosley06,Gehrels09,Soderberg10}). The precise
nature of the central engine which powers GRB-SNe, however, remains
an open question.

In this paper, we have undertaken extensive broadband ($>$100 MeV,
radio, optical, and X-ray) observations of three long-duration GRBs
(GRB 080916C, GRB 090902B, GRB 090926A) detected by the LAT
instrument on the Fermi satellite. The bulk Lorentz factors  imply
that the fireball is ultra-relativistic, which indicates that
hyper-energetic bursts carry as high as $E_{iso}\sim 10^{54}$ ergs
in the blast wave. The temporal indices $\alpha$ and spectral
indices $\beta$ in Tab.~\ref{tab:data}, which are consistent with
the observed simple power law slopes, show that our detailed density
transition consideration can reconcile the plateau and shallower
decay in the later afterglow. Tab.~\ref{tab:fitting} displays the
estimated physical parameter space of the best fit, which is all
reasonably certificated by the theory mentioned in
Sec.\ref{sect:fit}. Specifically, the wind parameter
$A_{35.5}\sim10^{\rm -2}-10^{\rm -1}$ shed us light on the central
engine of stellar collapsars and their association with SNe Ib/c
(\citealt{MacFadyen99,Heger06,Modjaz08}).

Furthermore, the tricky yet popular early rise of the LAT
($>$100MeV) light curve may indicate whether high energy photons
share the same source as the low energy afterglows. Aside from two
main kinds of explanation we mentioned above, leptonic or baryonic
models, an alternative possibility comes from the high-latitude
prompt emission \citep{Kumar00} and thus the flux density evolves as
$F_{\nu}(t)\propto t^{-2-\beta}$. However, according to
$\beta_{ob,LAT}$ in Tab.~\ref{tab:data}, the calculated temporal
index is too steep to explain the early high energy emission decay.
Or else if the GeV component results from the same conventional
unstructured ejecta as the late X-ray and optical afterglows, before
the deceleration time ($t<t_{dec}$, that is, we only consider the
thin shell case for long bursts), in the early wind bubble $n\propto
R^{-2}$, the bulk Lorentz factor is nearly constant, and thus the
electronic cooling and minimum frequencies evolve as $\nu_c\propto
t$ and $\nu_{\min}\propto t^{-1}$, the peak flux density
$F_{\max}\propto t^0$. Even when the SSC makes a significant effect,
the IC electronic cooling and minimum frequencies and IC peak flux
density evolve as $\nu_{IC,c}\propto t^{3}$ and $\nu_{IC,m}\propto
t^{-1}$ and $F_{IC,\max}\propto t^{-1}$. Since ${\rm
max}(\nu_{IC,c},\nu_{IC,m})>\nu_{GeV}>{\rm max}(\nu_{c},\nu_m)$
around the peak of LAT light curve, $F_{syn,GeV}\propto t^{(2-p)/2}$
(for $3>p>2$, this slope is not enough), as well as
$F_{IC,GeV}\propto t^{-2/3}$ or $t^{-(p+1)/3}$ for slow cooling and
$F_{IC,GeV}\propto t^{-2}$ or $t^{-5/2}$ for fast cooling. No matter
whether take the KN cutoff into account, this might be inconsistent
with the observed rise of the LAT light curve.

Our explanation for the initial rising part is simply based on the
assumption that $\Gamma\propto {t}^{\varkappa}$. Therefore,
$\nu_c\propto t^{{\varkappa}+1}$, $\nu_m\propto t^{{-\varkappa}-1}$,
and $F_{syn,GeV}\propto e^{-\tau_{\rm \gamma\gamma}}
t^{(1+\varkappa)(2-p)/2+2\varkappa} $ when $\nu_{GeV}>{\rm
max}(\nu_m,\nu_c)$. Co-contribution of the optical depth and
structured ejecta bring about the rising. This assumption can be
understandable within the framework of of the core collapse of a
massive star (i.e., SNe Ib/c). If the central engine ejects matter
continuously, a clean tunnel would be left after the early ejecta
sweeps up the stellar envelope, and the succeeding ejecta, whose
energy is attained via continuous accretion even after the LAT
trigger and whose Lorentz factor is larger, catches up with the
early ejecta and injects more energy into the blast wave, giving
rise to a rise of the LAT light curve. On the other hand, the
thermal component can also be explained under such assumption that
the previous jet lost kinetic energy due to impedance from shock
frontier, then turned it into thermal radiation. This hypothesis
have been provided by \citep{MacFadyen01} that, in the case of
red-giant or blue super giants , the powerful jet lose its energy
input at its base in the course of overtaking the sub-relativistic
weak supernova shock before reaching the surface, and eject little
highly relativistic matter; While in a helium star, the jet breaking
out with continuously accretion power receiving at the base will
lead its motion become highly relativistic. Later, some other
possibilities, such as \citet{Zhang03}'s baryon pollution scenario
and \citet{Fan09}'s initial envelope choked model, also result in
the earlier power-law rising distribution of bulk Lorentz factor.

In our frame, sub-MeV photons provide the background annihilation
with the later GeV ones. On one hand, photon with the largest energy
from the external shock fasten the constraint of lower Lorentz
factor of the ejecta [in \citet{Granot10}, the lower limit of the
bulk Lorentz factor should be $>1000$]; on the other hand, even the
ejecta is slower than the lower limit, it is possible that the
earlier rising is paretically contributed from opacity effect.

About the observed time delay of the pulses and variabilities in
both LAT and GBM bands, some other modulation effects, except for
optical depth can account for the concurrent variabilities in both
bands, such as inhomogeneous surroundings, energy injection, etc..
These factors indicate that the variability connection between LAT
and GBM \citep{Bzhang10} cannot give a sufficient clue to the
internal/external origin of the GeV photons.

In particular, as for GRB090902B, one of the tricky properties is
its soft-hard-soft quasi-thermal spectrum evolution, since the
soft-hard evolution appears within $T_{\rm 90}$, it can be better
understood by our model that the initial outflow was dissipated as
thermal component by the denser shock front/photosphere, when it
breaks out of the envelope, the consequent highly relativistic
ejecta in a clean tunnel can be dissipated strong enough to produce
energetic non-thermal emission, even the photosphere can be
outshined to some extent, therefore, the spectrum evolves from soft
to hard.

\begin{acknowledgements}
 This work is supported by the National Natural
Science Foundation of China (grants 10873009 and 11033002) and the
National Basic Research Program of China (973 program) No.
2007CB815404.
\end{acknowledgements}

\newpage


\begin{landscape}
\begin{table}

\begin{center}
\begin{tabular}{c|c c|c c|c c c c|c c c}\hline\hline
GRBs &$\alpha_{LAT}$ &$\alpha_{ob,LAT}$
 &$\beta_{LAT}$ &$\beta_{ob,LAT}$
 &$\alpha_{x,o,1}$   &$\alpha_{x,o,2}$
 &$\alpha_{ob,x}$ &$\alpha_{ob,o}$  &$\beta_{x,o}$ &$\beta_{ob,x}$ &$\beta_{ob,o}$
\\
\hline $080916C^{a}$&$1.27$ &$1.33\pm0.08$ &$1.16$ &$1.1\pm0.1$
 &$1.49$  &$0.99$
  &$1.29\pm0.09$ &$1.40\pm0.05$  &$0.66$ &$0.50\pm0.16$  &$0.38\pm0.02$
\\
 $090902B^{b}$ &$1.33$ &$1.40\pm0.06$ &$1.2$
&$-^{\clubsuit}$
 &$1.49$  &$0.98$
 &$1.36\pm0.03$ &$0.89\pm0.05$  &$0.7$
 &$0.90\pm0.13$ &$0.76\pm0.07$
\\
  $090926A^{c}$ &$2.01$ &$2.05\pm0.14$ &$1.5$
&$1.26_{-0.24} ^{+0.22}$
 &$2.00$  &$1.50$
  &$1.43\pm0.03^{\star}$ &$1.38\pm0.02^{\star}$  &$1.00$
  &$1.12\pm0.13^{\star}$ &$1.03\pm0.05^{\star}$
\\
\hline
 \hline
\end{tabular}
\caption{According to $F_{\nu}\propto t^{-\alpha}\nu^{-\beta}$,
$\alpha_{LAT}$, $\alpha_{x,o,1}$, $\alpha_{x,o,2}$ are the estimated
temporal indices of the LAT band photons, X-ray + optical photons in
the wind bubble, and X-ray + optical photons in the homogeneous
medium from our best-fit parameters. $\alpha_{ob,LAT}$,
$\alpha_{ob,x}$, $\alpha_{ob,o}$ are the observed temporal indices
of GeV photons, simple power-law X-ray photons and optical photons.
$\beta_{LAT}$ and $\beta_{x,o}$ are the estimated spectral indices
of the LAT band photons and X-ray + optical photons in either medium
from our best-fit parameters. $\beta_{ob,LAT}$, $\beta_{ob,x}$, and
$\beta_{ob,o}$ are the observed spectral indices of GeV photons,
simple power-law X-ray photons and optical photons. High energy data
are taken from \citet{Bzhang10}, X-ray data are taken from
\textit{Swift} Group (\citealt{Evans07,Evans09}). In addition, the
data labeled with superscript $a$ are taken from \citet{Greiner09},
the data with $b$ are from \citet{Pandey10}, the data with $c$ are
from \citet{Swenson10}, $-^{\clubsuit}$ represents no data
available. For 090926A, the data labeled with $\star$ are from
\citet{Cenko10}, the temporal slope of the X-ray and optical band in
\citet{Rau10} is: $\alpha=1.6\pm0.2$ before the first flare,
$\alpha=1.75\pm0.04$ after the second flare, and
$\alpha=1.63\pm0.01$ between, which is consistent with our estimated
value.}\label{tab:data}

\end{center}

\end{table}



\begin{table}

\begin{center}
\begin{tabular}{c| c c c c c c c c c c c|c c c}\hline\hline
GRBs       &$p$   &$\varkappa$   &$E_{K,iso,55}$ &$\eta_{3}$
&$A_{\star,35.5}$ &$\epsilon_{e}$ &$\epsilon_{B}$
&$\xi$ &$t_{break}$(s)  &$t_{end}$(s) &$t_{jet}$(s)&$\varsigma_{\gamma}$  &$R_{break}$(cm) &$n_{after}(cm^{-3})$\\
\hline $080916C$   &2.32   &0.5   &0.13  &0.74
&$0.02$ &0.3  &$5.81\times10^{-6}$ &2.1 &$10^{5.3}$  &$10^{5.6}$ &$-$ &$87.1\%$ &$3.95\times10^{18}$ &$1.17\times 10^{-3}$\\
$090902B$   &2.4   &0.8  &0.44  &0.60
&$0.06$ &0.4  &$4.56\times10^{-7}$ &2.0 &$10^{5.1}$  &$10^{6.0}$ &$10^{6.0}$ &$46.4\%$ &$7.52\times10^{18}$ &$1.62\times 10^{-3}$\\
090926A     &3.0   &0.3   &0.43  &0.56
&$0.03$ &0.35  &$8.18\times10^{-6}$ &2.5 &$10^{4.8}$  &$10^{4.9}$ &$-$&$33.0\%$ &$4.67\times10^{18}$ &$1.31\times 10^{-3}$\\

\hline
 \hline
\end{tabular}
\caption{The left part of the table are best-fit parameters for
GRB080916C, GRB090902B, and GRB090926A, the right part are prompt
emission energy fraction, radius when a density jump occurs and
circumburst density after the jump,
respectively.}\label{tab:fitting}
\end{center}

\end{table}

\end{landscape}

%

\begin{figure}
\includegraphics[width=1.0\textwidth,clip=]{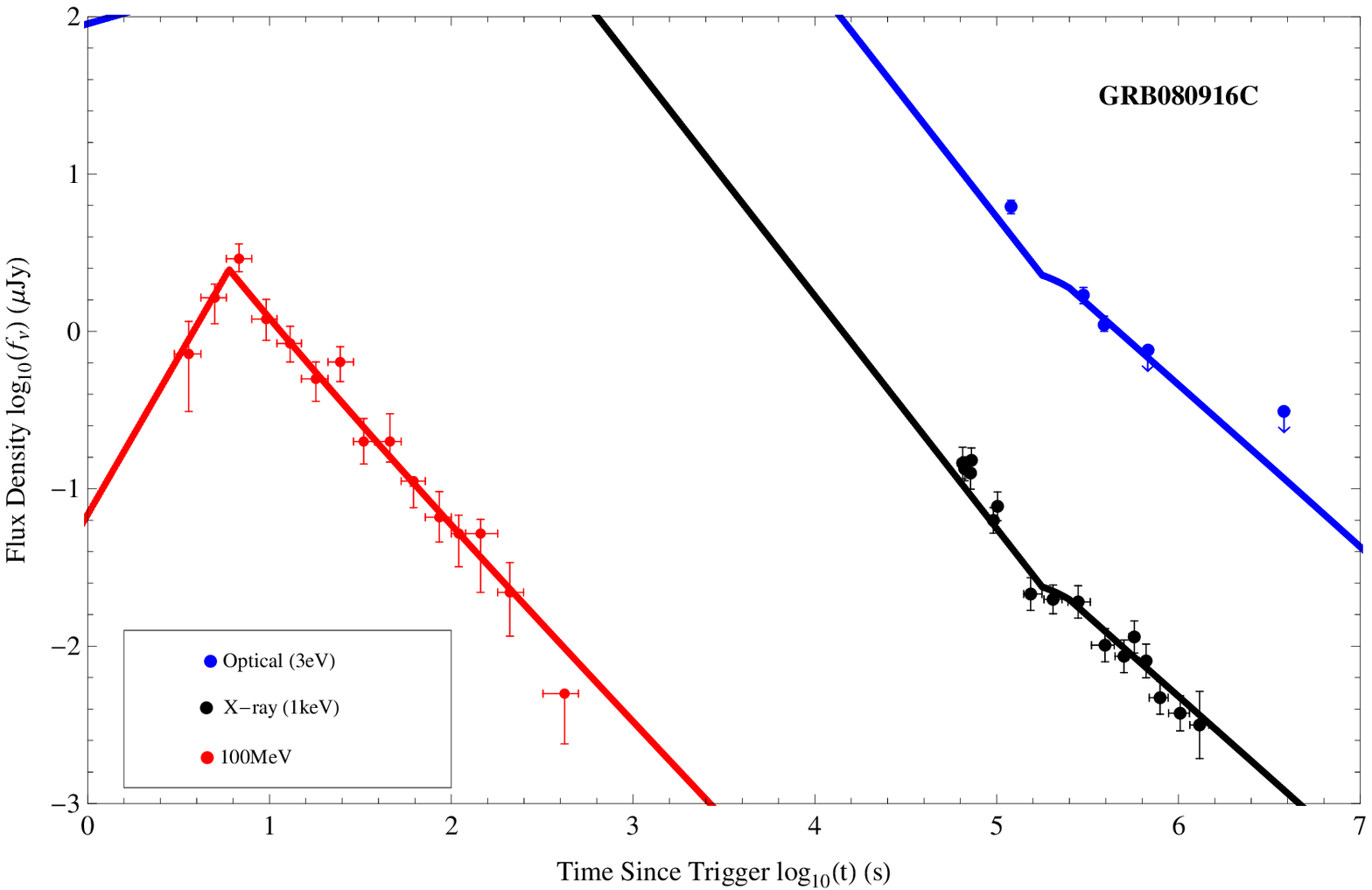}
\includegraphics[width=1.0\textwidth,clip=]{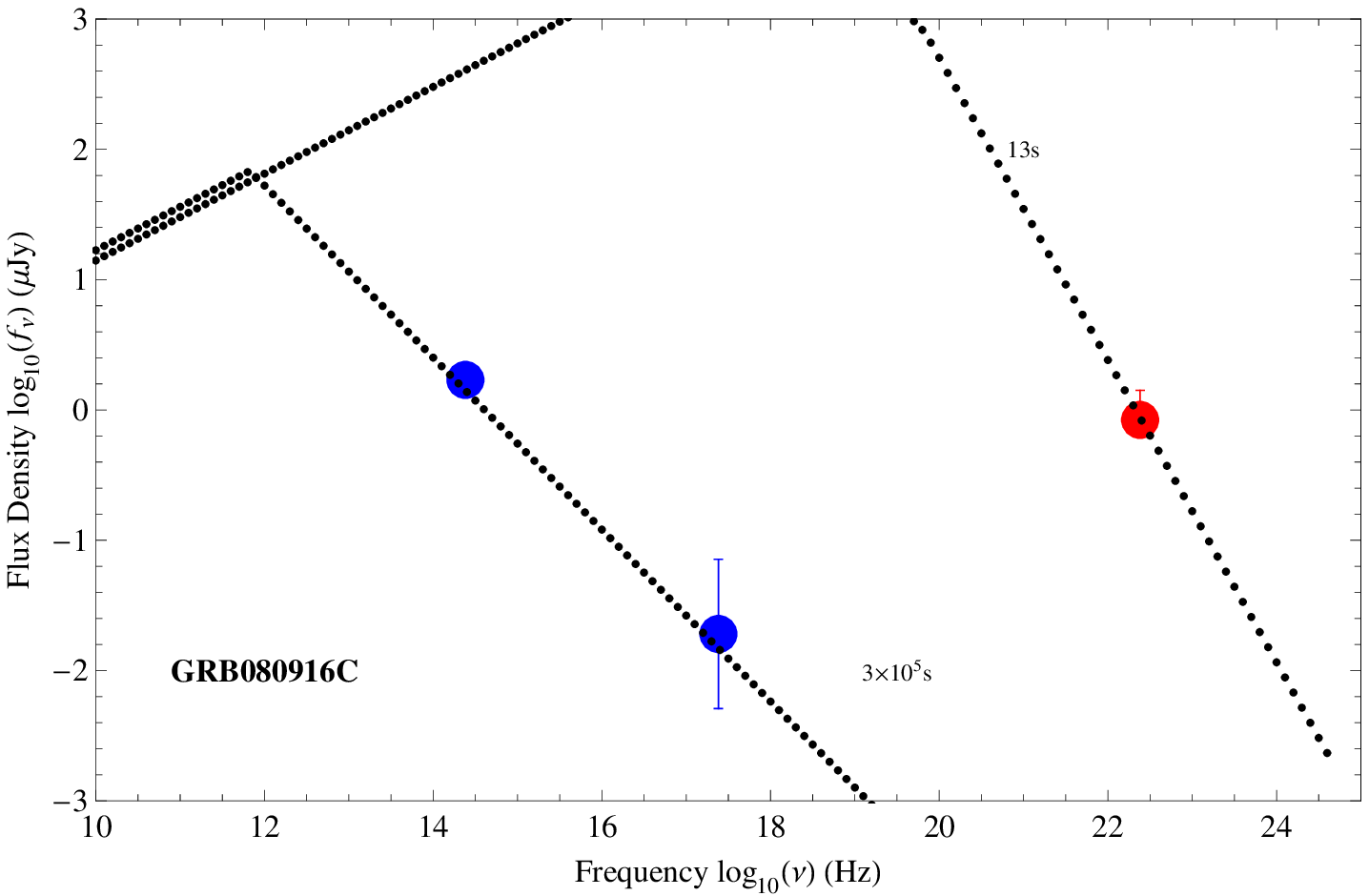}
\caption{Best fit for GRB 080916C: $p=2.32$, $\varkappa=0.5$,
$E_{K,55}^{iso}=0.13$, $\eta_3=0.74$, $A_{35.5}=0.02$,
$\epsilon_{e,0}=0.3$, $\epsilon_{B,0}=5.81\times10^{-6}$,
$\varepsilon=2.1$, $t_{break}=10^{5.3}$ s, $t_{end}=10^{5.6}$ s, no
jet break appears during the modeling period}\label{fig:080916C}
\end{figure}

\begin{figure}
\includegraphics[width=1.0\textwidth,clip=]{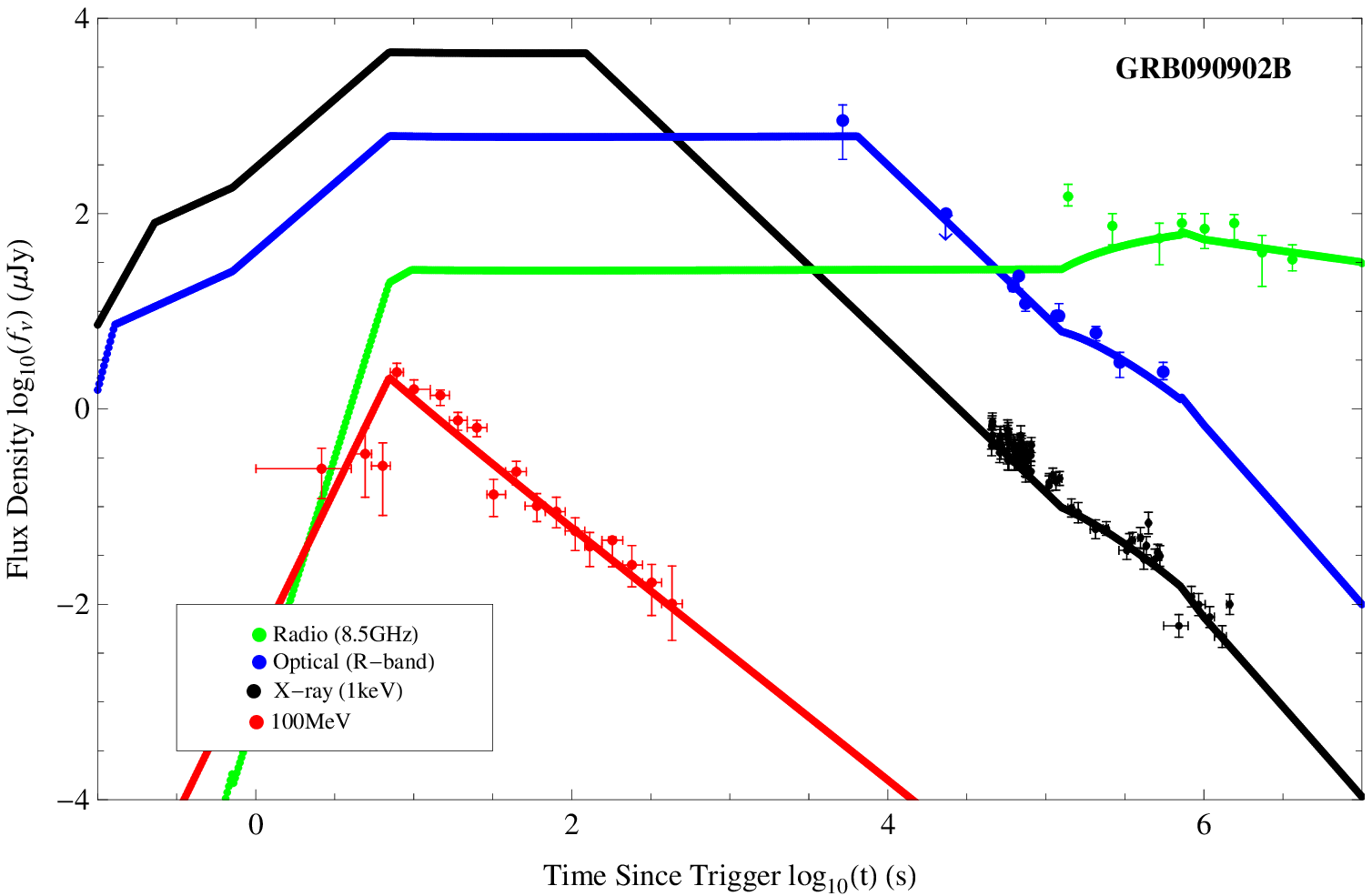}
\includegraphics[width=1.0\textwidth,clip=]{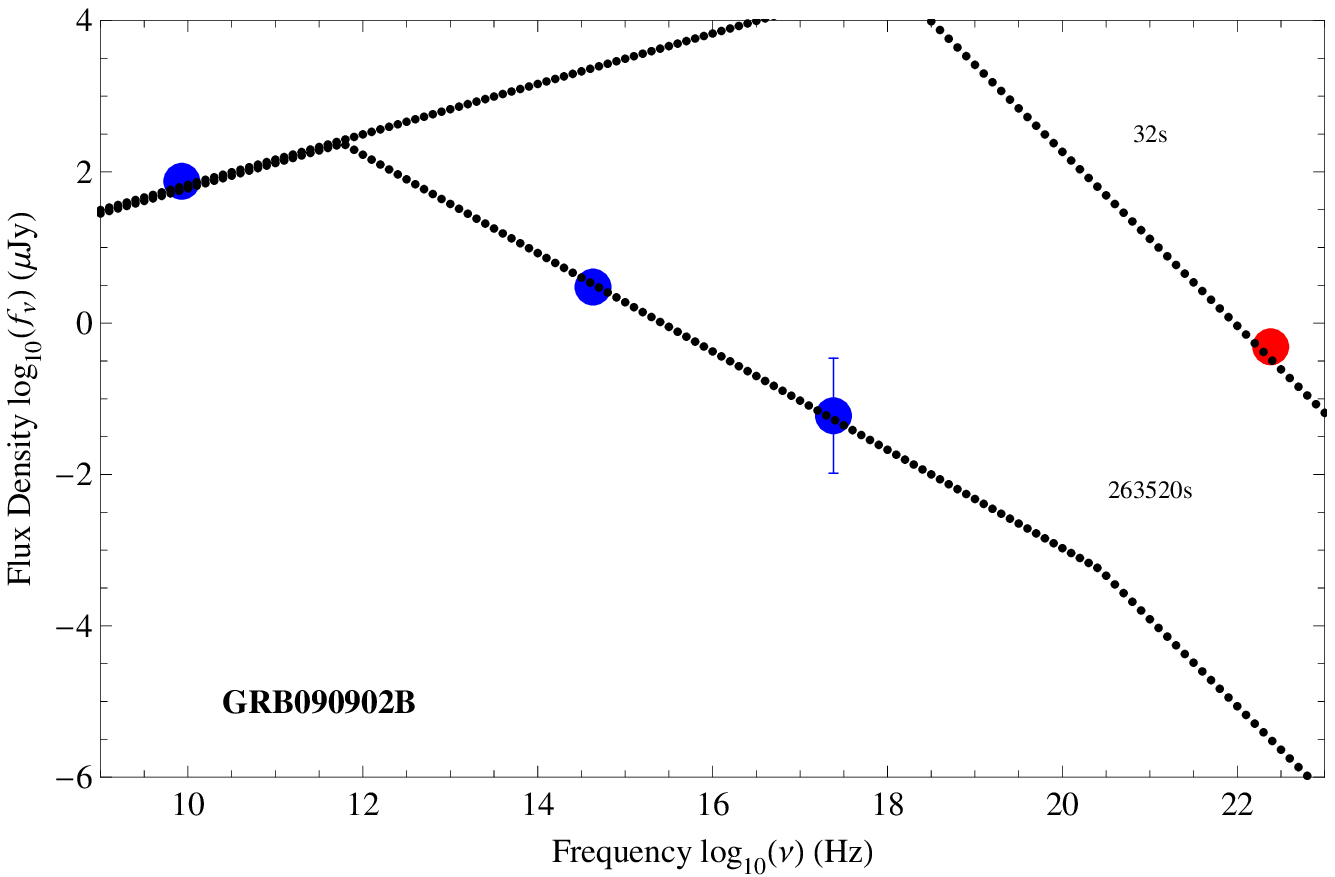}
\caption{Best fit for GRB 090902B: $p=2.4$, $\varkappa=0.8$,
$E_{K,55}^{iso}=0.44$, $\eta_3=0.60$, $A_{35.5}=0.06$,
$\epsilon_{e,0}=0.4$, $\epsilon_{B,0}=4.56\times 10^{-7}$,
$\varepsilon=2.0$, $t_{break}=10^{5.1}$ s, $t_{end}=10^{6.0}$ s,
$t_{jet}=10^{6.0}$ s}\label{fig:090902B}
\end{figure}

\begin{figure}
\includegraphics[width=1.0\textwidth,clip=]{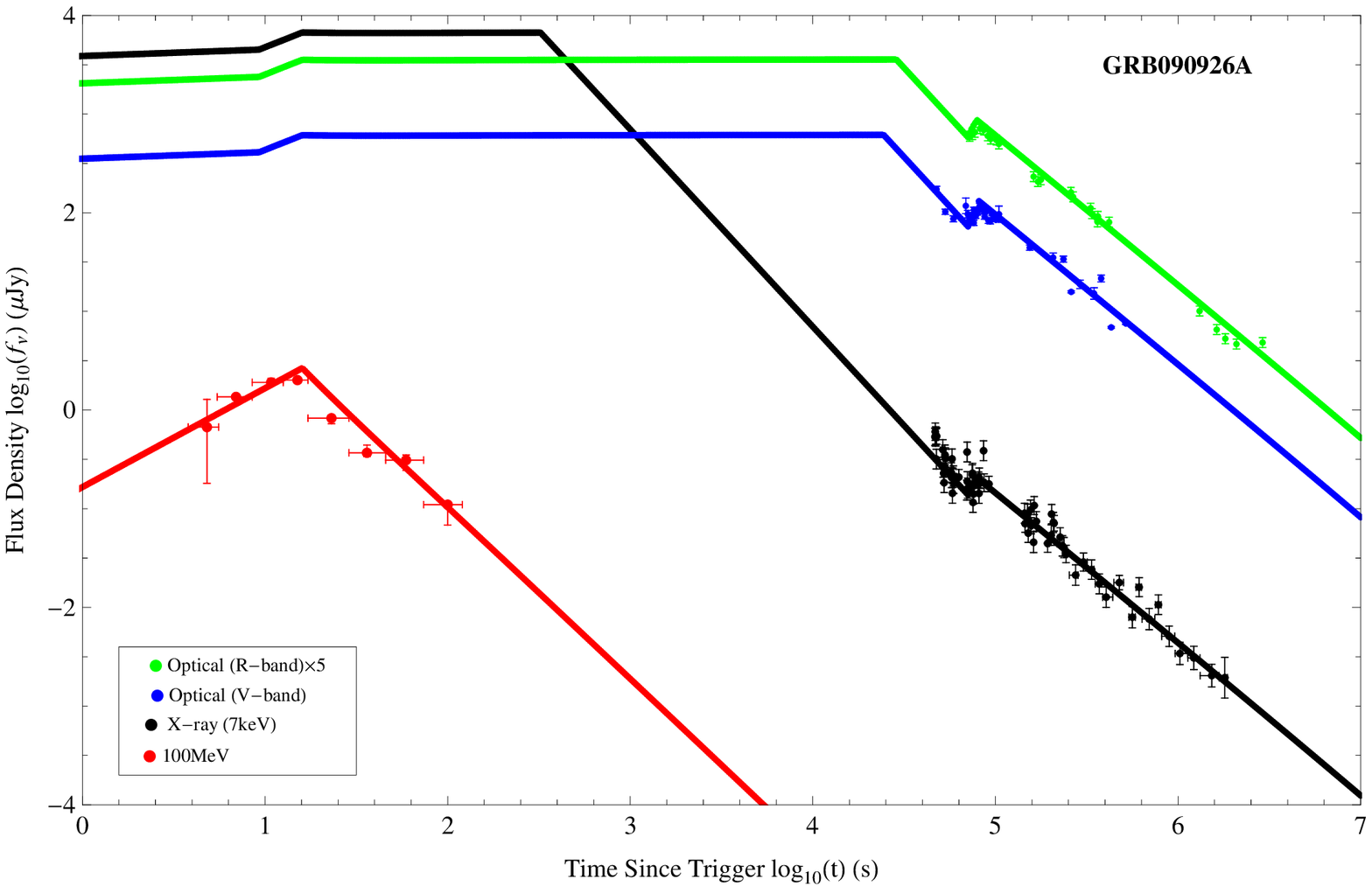}
\includegraphics[width=1.0\textwidth,clip=]{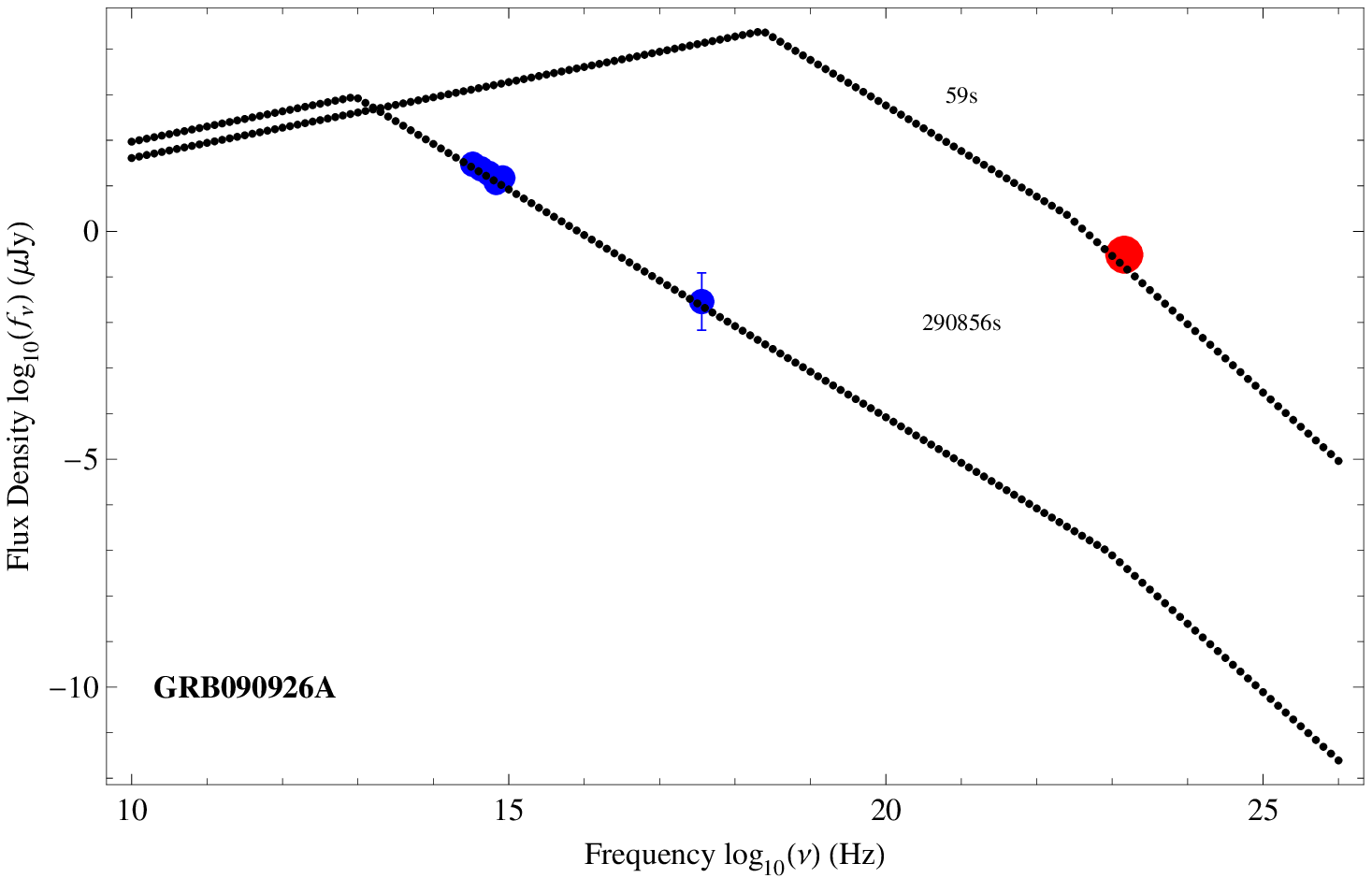}
\caption{Best fit for GRB 090926A: $p=3.0$, $\varkappa=0.3$,
$E_{K,55}^{iso}=0.43$, $\eta_3=0.56$, $A_{35.5}=0.03$,
$\epsilon_{e,0}=0.35$, $\epsilon_{B,0}=8.18\times10^{-6}$,
$\varepsilon=2.5$, $t_{break}=10^{4.8}$ s, $t_{end}=10^{4.9}$ s, no
jet break confirmed until 21 days}\label{fig:090926A}
\end{figure}

\begin{figure}
               \includegraphics[width=0.5\textwidth,clip=]{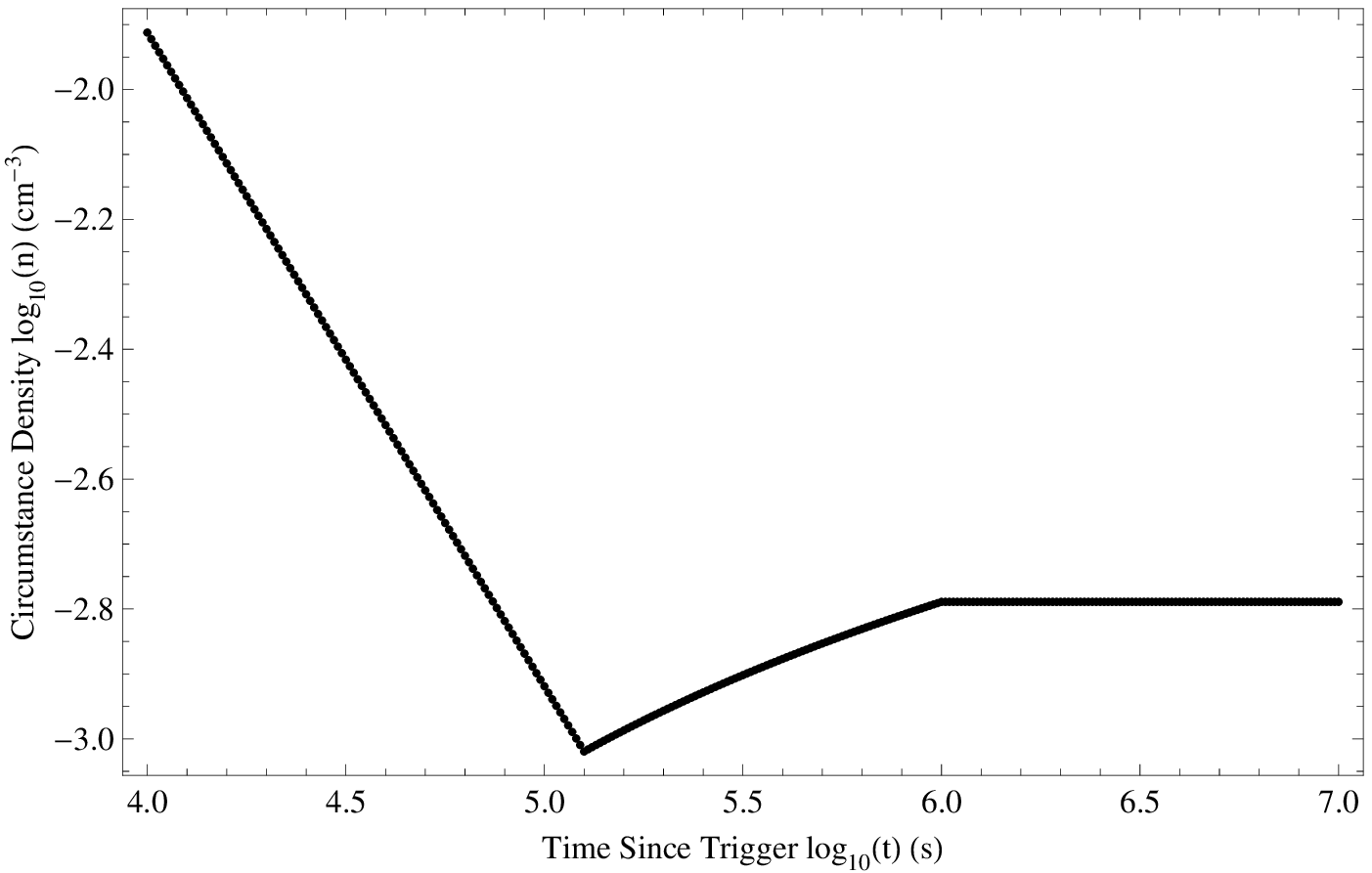}
               \includegraphics[width=0.5\textwidth,clip=]{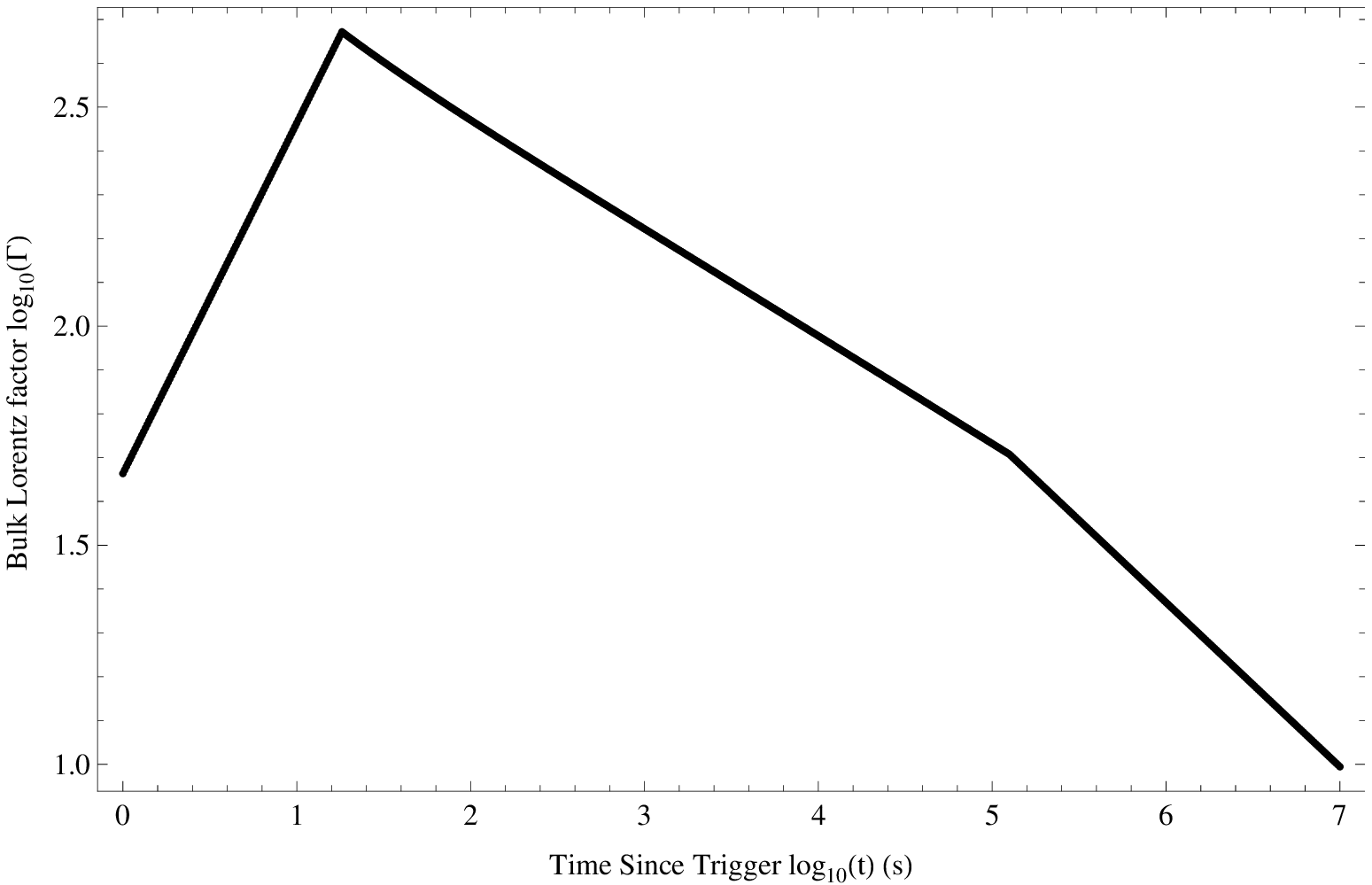}
               \includegraphics[width=0.5\textwidth,clip=]{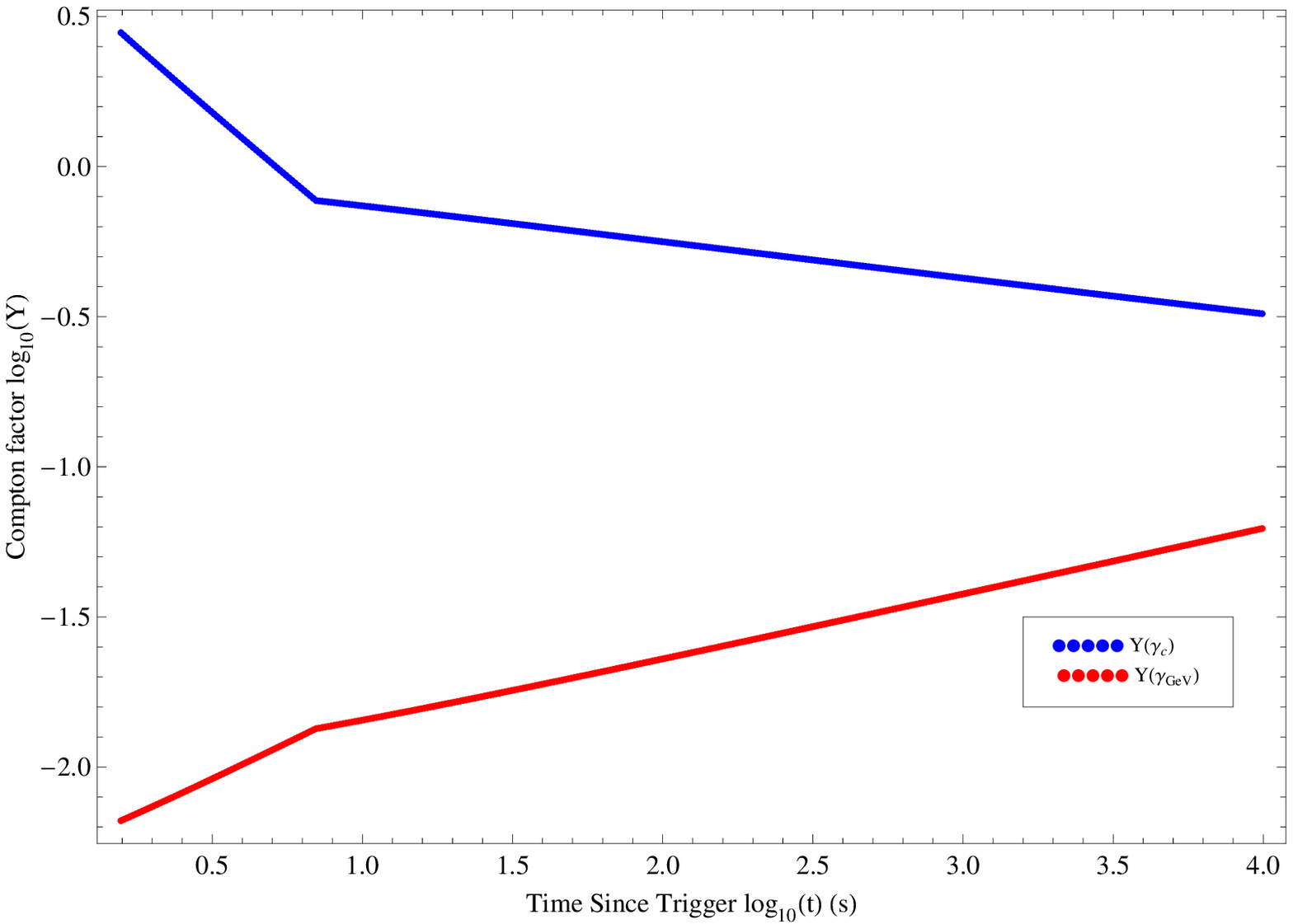}
               \includegraphics[width=0.5\textwidth,clip=]{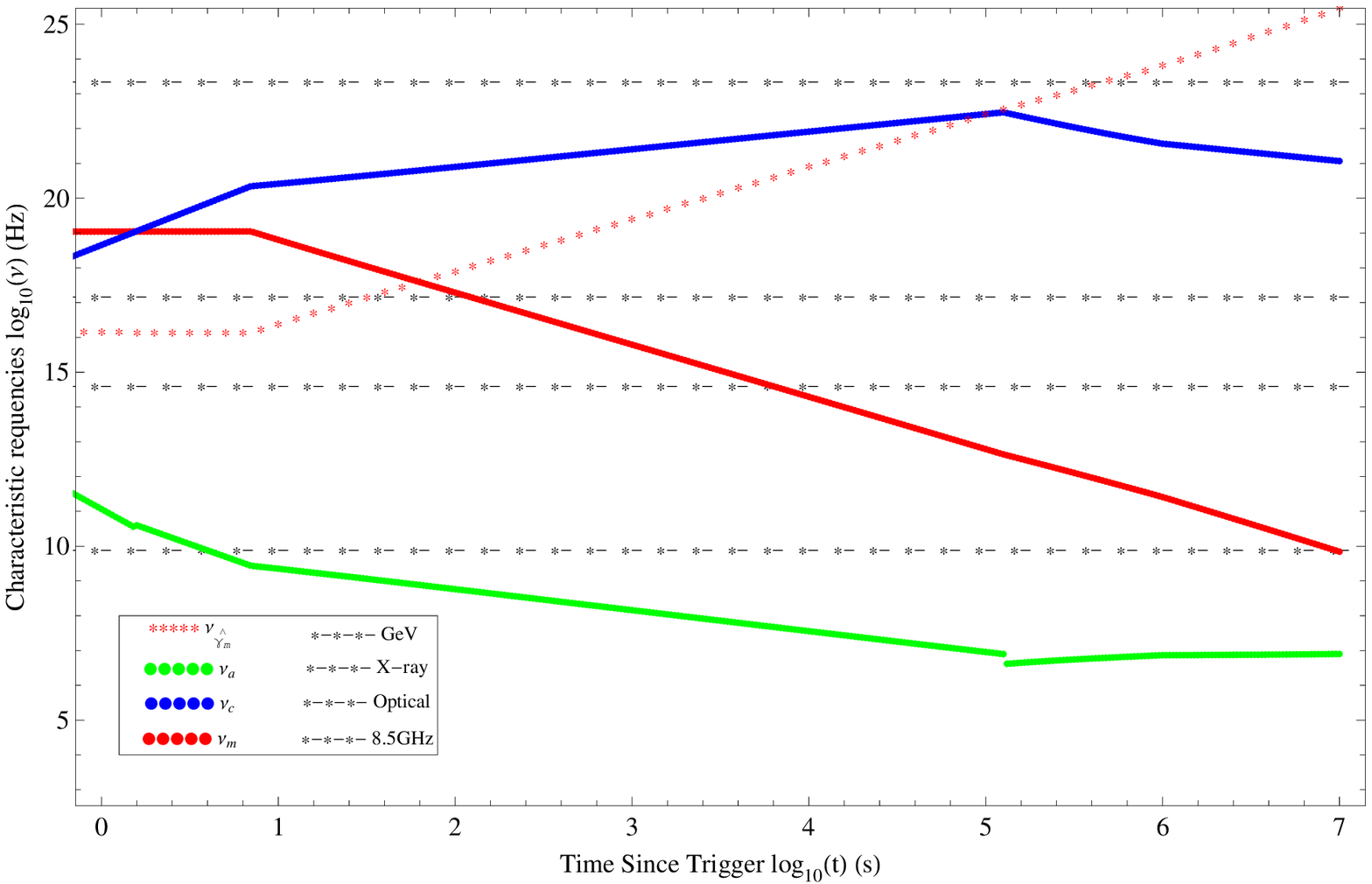}
\caption{Physical parameter evolution with the best fit parameter
for GRB090902B, similar to the other two bursts. Upper-left panel:
a. circumburst density with radius, upper-right panel: b. blast wave
bulk Lorentz factor with time; lower-left panel: c. Compton $Y$
factor ($\ll$1 due to KN effects) with time; lower-right panel: d.
characteristic frequency evolution.}\label{fig:parameter}
\end{figure}
%
\label{lastpage}

\end{document}